\documentclass[11pt,a4paper]{article}
\usepackage{float}
\usepackage{jheppub}
\makeatletter
\def\@fpheader{}
\makeatother
\usepackage[T1]{fontenc}
\usepackage{lmodern,amsmath,amssymb,booktabs,microtype,placeins,tikz}
\usepackage{ragged2e}
\usepackage{sidecap}
\usepackage{float}
\usetikzlibrary{arrows.meta}
\hypersetup{pdftitle={Mixed-sign QNM Landau cut and real-time response},pdfauthor={}}

\newcommand{\qn}{{\mathfrak{q}}}
\newcommand{\wn}{{\mathfrak{w}}}
\usepackage{bm}

\newcommand{\vextmom}{\bm{\kappa}}

\begin{document}

\title{From Quasinormal Mode Poles to Branch Cuts at Finite $N$}

\author{%
	{\large\bfseries Navid Abbasi}\\[3pt]
	\normalsize School of Nuclear Science and Technology, Lanzhou University, China\\
	\small\texttt{abbasi@lzu.edu.cn}%
}

\abstract{
    In the strict large-$N$ limit, holographic thermal retarded correlators are typigcally meromorphic, with their late-time behavior determined by isolated quasinormal mode (QNM) poles. We show that the first bulk-loop correction, of order $1/N^2$, qualitatively changes this structure. In a scalar theory on the planar AdS$_5$ black brane, a one-loop bulk diagram generates poles at sums of two QNM frequencies at fixed loop momentum $q$; integration over the continuous momentum turns these QNM-pair poles into branch cuts. At vanishing external spatial momentum, for pairs built from QNM branches with opposite real parts, the leading light-cone contributions cancel at large momentum, and the residual QNM-pair frequency scales as $q^{-1/3}$. Consequently, an infinite family of cuts accumulates at zero frequency, while all tree-level QNM poles remain at finite nonzero frequencies. The first finite-$N$ correction therefore closes the gap set by the classical QNM spectrum. Competition between the decreasing QNM-pair damping and thermal Boltzmann suppression produces a moving saddle and a finite-$N$ contribution $G(t)\sim N^{-2}t^\alpha e^{-Ct^{3/4}},\, C>0,$ which decays more slowly than the leading classical QNM contribution. Thus the large-$N$ expansion is nonuniform at late times, and bulk quantum fluctuations generate a distinct  late-time regime beyond the classical QNM decay.
}

\maketitle

\section{Introduction and review of the results}
\label{sec:intro}
The late-time decay of small perturbations around thermal equilibrium is controlled by the analytic structure of retarded correlation functions. In holography, this structure has been understood predominantly in the strict large-\(N\) limit, where the dual bulk description reduces to classical gravity.  In that limit the response is
organized by the quasinormal modes (QNMs) of the black brane: ingoing
normalizable bulk fluctuations appear as poles of boundary retarded
correlators, and their imaginary parts set the characteristic damping scales
of linear response
\cite{	Horowitz:1999jd,Birmingham:2001pj,Son:2002sd,Nunez:2003eq,Kovtun:2005ev,Berti:2009kk,Festuccia:2008zx} \footnote{Quasinormal modes have since become a standard tool for characterizing near-equilibrium holographic dynamics
	\cite{Kaminski:2009dh,Ammon:2017ded,Baggioli:2020magnetophonons,
		Davison:2018ofp,Meiring:2023wwi,Wang:2021holo,Ge:2015uaa,Mamani:2022qnm,Chen:2025glueball,Li:2019dgl,He:2023thermal,Tan:2023brane,Finazzo:2016psx,Gursoy:2016ggq}.
		Their continuation to complex frequency and momentum has revealed mode collisions, branch points, and connections between hydrodynamic and nonhydrodynamic sectors
		\cite{Withers:2018srf,Grozdanov:2019uhi,Grozdanov:2019kge,
			Abbasi:2020ykq,Heller:2020uuy,Arean:2023ejh}. Analytically controlled limits such as large $D$ have likewise revealed universal QNM sectors with parametrically weak damping \cite{Emparan:2014cia}.
		For broader perspectives, see Refs.~\cite{Heller:2016rtz,Florkowski:2017olj}. Recent work has also shown that smooth deformations can qualitatively reorganize QNM spectra, including the emergence of nonzero real parts from initially purely imaginary modes \cite{Jia:2026ncd}.}.
Beyond the strict large-$N$ limit, however,  interactions among bulk fluctuations and
the associated quantum effects can qualitatively modify this analytic
structure.  This leads to the questions that motivate this work:
\emph{what analytic structure supplements the classical QNM poles once such
	finite-$N$ effects are included, and which part of that structure controls
	the late-time response?}

Recent work has substantially sharpened the large-\(N\) picture. Exact and semiclassical thermal correlators and their analytic structure \cite{Bhatta:2022wga,Jia:2024qnm,Buric:2025kms,Arnaudo:2026fourier,Jia:2026spectral}, thermal product representations \cite{Dodelson:2023vrw,Bhattacharya:2025shear}, and KMS and spectral-duality constraints \cite{Grozdanov:2024spectral,Grozdanov:2026spectral} have made increasingly explicit how QNM spectra encode the analytic structure of holographic response. 
More broadly, studies of holographic retarded correlators at complex frequency and momentum have revealed nontrivial structures and constraints on their pole trajectories, including relations to near-horizon bulk dynamics \cite{Blake:2018leo,Blake:2019otz}.
Related reconstruction methods have shown that, in suitable cases, substantial information about the QNM spectrum and correlator can be recovered from pole-skipping data \cite{Grozdanov:2023tag}. 
Closely related studies have connected singularities of thermal correlators to bulk causal structure, OPE data, and black-hole singularities \cite{Dodelson:2024bulkcone,Ceplak:2024ope,
	AfkhamiJeddi:2026singularity,Grozdanov:2026bouncing}.
Nonanalytic structures can also emerge in special classical limits: in near-extremal backgrounds, nonhydrodynamic poles can condense into a branch cut as \(T\to0\) \cite{Gouteraux:2025kta}.
Most of these developments concern the leading large-\(N\), classical-gravity regime. What remains much less understood is the new analytic structure generated by the first bulk loop.

It is useful to distinguish the coupling and $1/N$ expansions. At leading order in $1/N$, finite-coupling effects are encoded by stringy or higher-derivative corrections to the bulk action and can modify the classical QNM spectrum already at tree level \cite{Stricker:2013lma,Waeber:2015oka,Grozdanov:2016zjj,Buchel:2018}. Finite-\(N\) effects instead organize the bulk loop expansion \cite{Aharony:1999ti,Aharony:2016dwx}. These are independent expansions: a higher-derivative bulk theory can itself receive quantum-loop corrections.

With this distinction in mind, weakly coupled finite-temperature field theory provides a useful point of comparison. There, loops of thermal quasiparticles with continuous relative momentum generate multiparticle branch cuts \cite{Weldon:1998tz}. At strong coupling, a quasiparticle description need not exist; in the classical holographic regime, thermal response is instead organized by isolated QNM poles, and the contrast between weak-coupling branch cuts and strong-coupling QNM poles has been emphasized in Refs.~\cite{Hartnoll:2005ju,Hou:2010st}. The question we address is how robust this pole-based description remains once the first bulk quantum correction is included.

A closely related holographic precedent is the one-loop origin of hydrodynamic long-time tails. Caron-Huot and Saremi showed that bulk loops reproduce the familiar power-law tail associated with hydrodynamic fluctuations \cite{CaronHuot:2009iq}; see also Refs.~\cite{Kovtun:2003vj,Kovtun:2011np,Kovtun:2012rj,An:2020vri,Basar:2025review}. This demonstrates that bulk quantum fluctuations can generate nonanalytic structure beyond the classical poles. Their analysis, however, concerns gapless hydrodynamic modes and the resulting singular behavior near \(\omega=0\).  Here we instead ask how a bulk loop built from gapped nonhydrodynamic QNMs modifies the analytic structure of a thermal retarded correlator.

To study this problem, we formulate the one-loop retarded correlator using the real-time Schwinger–Keldysh formalism. Thermal bulk propagators and general real-time gauge/gravity prescriptions were established  Refs.~\cite{Herzog:2002pc,Skenderis:2008dh,Skenderis:2008dg}, followed by holographic Schwinger–Keldysh effective theories and open-system formulations \cite{deBoer:2018qqm,Jana:2020vyx,Loganayagam:2022zmq}. Thermal higher-point functions were constructed directly from Schwinger–Keldysh contours in Ref.~\cite{Pantelidou:2022ftm}. More recently, an exterior effective field theory description of real time black hole dynamics was developed \cite{Loganayagam:2025exterior} and extended to bulk loop diagrams  \cite{Loganayagam:2026loops}.
These developments provide the framework for computing real-time bulk quantum corrections. We use this framework below to determine the analytic structure generated by the QNM spectrum at first loop order.

\begin{figure}[t]
	\centering\includegraphics[width=0.8\textwidth]{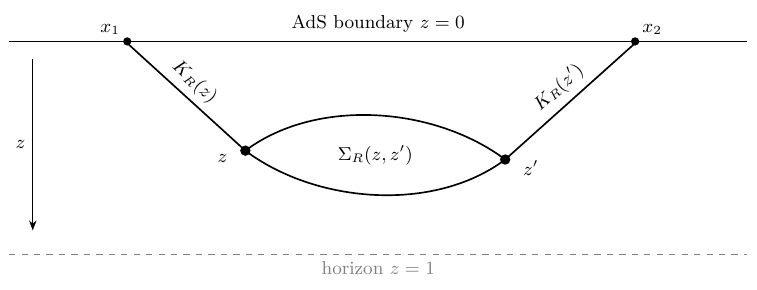}
	\caption{Schematic bulk one-loop contribution to the boundary retarded correlator.
		The external retarded bulk-to-boundary propagators $K_R(z)$ and $K_R(z')$
		connect the boundary insertions to bulk vertices at radial positions
		$z$ and $z'$, while $\Sigma_R(z,z')$ denotes the retarded bulk self-energy
		kernel generated by the loop. The two radial positions are integrated
		independently and are drawn at similar depths only for visual clarity.}
	\label{intro_loop}
\end{figure}

A direct calculation of this problem is considerably harder than the classical QNM problem. Let us suppose that the bulk theory contains interactions capable of generating the one-loop correction shown schematically in Fig.~\ref{intro_loop}. Two external retarded legs connect the boundary insertions to bulk interaction points at radial positions \(z\) and \(z'\), while two real-time bulk propagators run around the loop. The calculation therefore involves integrations over the loop frequency and momentum, together with independent integrations over the two bulk radial positions. The quantity of interest is the resulting boundary correlator.

The singular part of the correlator can be isolated more efficiently. At fixed loop momentum, the two internal propagators have QNM poles in the complex loop-frequency plane. As the external frequency is analytically continued, a pole from each propagator can approach the loop-frequency contour from opposite sides. When the two poles meet, the contour becomes pinched and the correlator develops a nonanalyticity. This is the thermal analogue of the familiar Landau-pinch mechanism in finite-temperature field theory
~\cite{Landau:1959fi,Weldon:1983jn,Weldon:1998tz}. The pinch produces a pole at the sum of the two QNM frequencies. The remaining momentum integration determines the resulting branch-cut structure, while the radial integrations determine whether the singular contribution survives in the boundary correlator and with what strength.

\begin{figure}[t]
\includegraphics[width=1.05\textwidth]{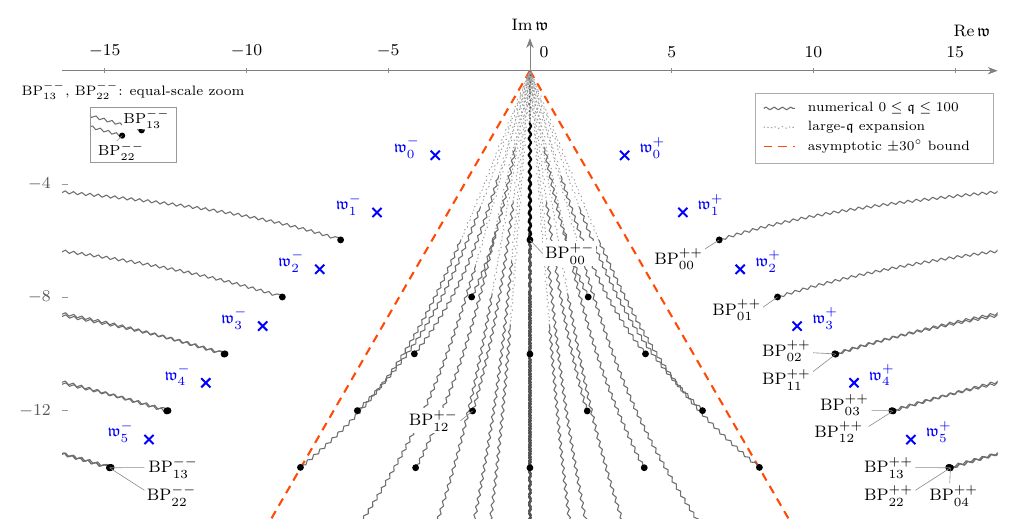}
	\caption{
		Representative QNM-pair branch cuts at vanishing external spatial momentum, \(\kappa=0\). Blue crosses denote the classical QNM frequencies \(\mathfrak w_n^\pm(0)\), while black dots mark the \(\qn=0\) endpoints of the branch cuts. The solid zigzags show the numerically determined cuts for \(0\le \qn\le100\), parameterized by the loop momentum \(\qn\); faint dotted zigzags show their large-\(\qn\) asymptotic continuation. The \(\qn=0\) endpoints are square-root branch points of the boundary correlator. The diagonal mixed-sign cuts lie on the negative imaginary axis, with the fundamental \(00^{+-}\) channel emphasized; off-diagonal mixed-sign cuts occur in reflected pairs and accumulate at the origin as \(\qn\to\infty\). The red dashed lines bound the large-\(\qn\) asymptotic sector, \(|\theta|<30^\circ\), derived in Eq.~\eqref{theta_m_n}. The top-left inset resolves the closely spaced same-sign branch points.
			}
	\label{intro_overview}
\end{figure}

For simplicity, we first specialize to vanishing external spatial momentum, \(\kappa=0\) \footnote{Throughout, frequencies and spatial momenta are measured in units of
	$\pi T$:
	\begin{equation}
		\mathfrak w\equiv\frac{\omega}{\pi T},
		\qquad
		\kappa \equiv\frac{k}{\pi T},
		\qquad
		\mathfrak q\equiv\frac{q}{\pi T},
		\qquad
		\tau\equiv\pi T\,t,
	\end{equation}
	where $\mathfrak \kappa$ denotes the external momentum and $\mathfrak q$ the
	loop momentum. The restriction to $\boldsymbol{\kappa}=0$ is made to expose the structure in its simplest form. We comemnt on  $\boldsymbol{\kappa}\ne0$ in Sec.~\ref{sec:discussion}.}. The two internal propagators then carry momenta \(\boldsymbol{\qn}\) and \(-\boldsymbol{\qn}\), with \(\qn=|\boldsymbol{\qn}|\). We denote their QNM frequencies by \(\wn_n^\pm(\qn)\), where \(n=0,1,\ldots\) labels the overtone and the signs distinguish the positive and negative real frequency branches. The possible pinches are organized by the pair frequencies
\begin{equation}
\Omega_{mn}^{\sigma\tau}(\qn)
=\wn_m^\sigma(\qn)+\wn_n^\tau(\qn),
\qquad \sigma,\tau=\pm .
\end{equation}
As the loop momentum varies, these pair frequencies trace curves in the complex-frequency plane.
These curves identify candidate branch cuts, but not every point on them is a branch point of the boundary correlator. After the momentum integration, branch points occur at special values of \(\qn\), such as endpoints of the integration domain or stationary points of \(\Omega_{mn}^{\sigma\tau}(\qn)\), provided the corresponding boundary coefficient is nonzero. Varying the signs and overtone indices organizes the resulting pair-frequency loci into the \textit{ QNM-pair networkc} shown in Fig.~\ref{intro_overview}.

	The two sign sectors have qualitatively different geometries. Same-sign
	pairs generate the outer families shown in the figure. 
		For mixed-sign pairs, the reflection relation
	$\mathfrak w_n^-(\mathfrak q)
	=
	-\big[\mathfrak w_n^+(\mathfrak q)\big]^*$
	gives
	$\Omega_{mn}^{+-}(\mathfrak q)
	=
	\mathfrak w_m^+(\mathfrak q)
	-
	\big[\mathfrak w_n^+(\mathfrak q)\big]^*.$
	The real parts therefore cancel for $m=n$, placing the diagonal mixed
	channels on the negative imaginary axis. For $m\neq n$, this cancellation
	is incomplete, and exchanging the two indices reflects the pair-frequency
	curve across that axis. The central network thus contains both overlapping
	diagonal channels and reflected off-diagonal families.
	
	A further organizing principle emerges at large loop momentum. For each fixed overtone, the large-$\qn$ QNM asymptotics derived by Festuccia and Liu \cite{Festuccia:2008zx} take the form 
\begin{equation}\label{omega_m_plus}
	\wn_n^+(q)
	=
	\qn+\frac{\lambda_n}{2}e^{-i\pi/3}\qn^{-1/3}
	+O(\qn^{-5/3}),
\end{equation}
	where \(0<\lambda_0<\lambda_1<\cdots\) are the positive eigenvalues defined in App.~\ref{app:largeq}. Using the reflected negative real frequency branch then gives, for a mixed-sign pair,
\begin{equation}
	\Omega_{mn}^{+-}(\qn)
	=
	\frac14
	\left[
	\lambda_m-\lambda_n
	-i\sqrt3(\lambda_m+\lambda_n)
	\right]\qn^{-1/3}
	+O(\qn^{-5/3}).
\end{equation}
	Consequently, every mixed-sign pair approaches \(\wn=0\) as \(\qn\to\infty\), so the origin is an accumulation point of the entire mixed-sign tower. The direction of approach is also constrained. If \(\theta_{mn}\) is measured from the negative imaginary axis,
\begin{equation}\label{theta_m_n}
	\tan\theta_{mn}
	=
	\frac{\lambda_m-\lambda_n}
	{\sqrt3(\lambda_m+\lambda_n)},
	\qquad
	|\theta_{mn}|<\frac{\pi}{6}.
\end{equation}
	Thus all fixed-index mixed-sign channels approach the origin within the \(60^\circ\) sector bounded by the red dashed rays in Fig.~\ref{intro_overview}.	
	
	The accumulation at zero frequency does not imply that all mixed-sign channels contribute equally to late-time relaxation. For every mixed-sign channel, approaching the origin requires increasing the loop momentum: this reduces the QNM-pair damping, but at the same time suppresses the thermal weight through the Boltzmann factor $e^{-\beta q}=e^{-\pi \qn}$. The late-time response is therefore governed by a competition between these two effects. Their balance selects a momentum scale that grows as $\tau^{3/4}$ and produces a stretched-exponential decay, slower than any ordinary exponential $e^{-\gamma\tau}$ with fixed $\gamma>0$. The quantitative suppression depends on the QNM pair, and the saddle-point analysis gives
\begin{equation}\label{G_mix}
	G_{mn}^{\rm mix}(\tau)
	\sim
	\epsilon A_{mn}\,
	\tau^{\nu/2+1/8}
	e^{-C_{mn}\tau^{3/4}}.
\end{equation}
	
	Here \(A_{mn}\) determines the strength of the contribution from the \((m,n)\) mixed-sign channel and is obtained from the QNM residues and radial integrations. This asymptotic form applies whenever \(A_{mn}\neq0\). In Eq.~\eqref{G_mix}, \(\operatorname{Re}C_{mn}\) is positive and controls the stretched-exponential suppression of each channel. We show below that the fundamental \(00^{+-}\) channel has the smallest \(\operatorname{Re}C_{mn}\), and therefore decays most slowly at asymptotically late times. Its special role comes from this ordering of decay rates, not from being the only cut that reaches zero frequency.
	
	To summarize, at leading order in the large-\(N\) expansion, the scalar correlator is meromorphic and all of its isolated QNM poles remain a finite distance from zero frequency. At the first finite-\(N\) correction, branch cuts appear and extend to zero frequency. Thus the analytic structure changes qualitatively already at order \(1/N^2\), closing the gap set by the classical QNM spectrum. As a result, there is a late-time regime in which the stretched-exponential decay generated by the one-loop branch-cut contribution dominates over the exponential decay governed by the leading classical QNM contribution.

	
	The rest of the paper is organized as follows. Section~\ref{sec:setup} introduces the holographic model and the one-loop retarded correlator. In Sec.~\ref{sec:signed-pairs} we derive the QNM-pair singularity condition and the corresponding boundary coefficient. We than analyze the lowest mixed-sign channel and the resulting branch-cut network. In Sec.~\ref{sec:origin-accumulation} we study the large-momentum accumulation of the mixed-sign cuts near \(\wn=0\), and in Sec.~\ref{sec:real-time} we determine the resulting late-time response. Section~\ref{sec:discussion} discusses the implications and possible extensions. Technical details and numerical checks are collected in the appendices.

	\section{Holographic setup and one-loop retarded correlator}
	\label{sec:setup}
	
	We now formulate the one-loop problem in real-time holography. We first specify the bulk model and the retarded and thermal propagators entering the loop, and then project the first bulk self-energy insertion onto the boundary correlator. This construction also makes explicit how the QNM poles of the internal propagators arise from the bulk radial problem and enter the loop dynamics.
	
	\subsection{Bulk model and linear propagators}
	\label{sec:linear-propagators}
	
We work on the planar AdS$_5$--Schwarzschild black brane, whose horizon
radius is related to the temperature by
$z_h=\frac{1}{\pi T}.$
We introduce dimensionless horizon coordinates
\[
z=\frac{z_{\rm phys}}{z_h},
\qquad
\tau=\frac{t_{\rm phys}}{z_h}
=\pi T\,t_{\rm phys},
\qquad
\mathbf{x}=\frac{\mathbf{x}_{\rm phys}}{z_h}.
\]
In these coordinates the horizon is at $z=1$, and $z_h$ has been scaled
out of the metric.  Setting the AdS radius to one, we obtain
\begin{equation}
	ds^2
	=
	\frac{1}{z^2}
	\left[
	-f(z)d\tau^2
	+d\mathbf{x}^2
	+\frac{dz^2}{f(z)}
	\right],
	\qquad
	f(z)=1-z^4.
	\label{eq:metric}
\end{equation}
The AdS boundary is at $z=0$.  Since the physical Euclidean time has
period $1/T=\pi z_h$, the dimensionless Euclidean coordinate $\tau_E$
has period $\pi$.
	
	On this background we consider a scalar field \(\phi\) with a cubic self-interaction,
\begin{equation}\label{action}
	S=-\mathcal A\int d^5x\sqrt{-g}\left[
	\frac12(\nabla\phi)^2+\frac12m^2\phi^2
	+\frac{g}{3!}\phi^3+\cdots
	\right].
\end{equation}
		We choose \(m^2=1\) as a representative positive scalar mass; this keeps the spectrum gapped and avoids special zero-mode or boundary-degeneracy effects. The ellipsis denotes additional local interactions and counterterms needed to complete and renormalize the model. In particular, we impose the tadpole renormalization condition such that the scalar one-point function vanishes and \(\phi=0\) remains the background about which the two-point function is computed. These local terms do not affect the nonlocal singular structure studied below.
	
		Moreover,	rescaling the field as
		$\phi_c=\sqrt{\mathcal A}\,\phi$
	removes \(\mathcal A\propto N^2\) from the quadratic action and gives the canonically normalized cubic interaction \(g\,\phi_c^3/(3!\sqrt{\mathcal A})\). In what follows we work with the canonically normalized field and suppress the subscript \(c\). The loop expansion parameter is therefore
	\begin{equation}
		\epsilon\equiv \frac{g^2}{\mathcal A}\sim N^{-2}.
	\end{equation}
	
	\subsection{From classical response to quantum fluctuations.}
	\label{symmetrized_correlator}
	At the classical holographic level, the thermal state is encoded in the black-brane geometry, while the retarded response is determined by the classical bulk equations with retarded boundary conditions. This is sufficient to determine the QNM spectrum and the classical retarded correlator; no separate specification of the quantum state of bulk fluctuations is required.
	
At one loop, this is no longer sufficient; one must also specify the
quantum state of the bulk fluctuations. We take this to be the
Hartle--Hawking state, the stationary KMS state at the Hawking
temperature, regular on the full Kruskal extension of the black brane
\cite{Herzog:2002pc,Skenderis:2008dh} \footnote{On the maximally extended two-sided AdS black-brane geometry, the Hartle–Hawking state can be viewed as the thermofield-double purification of the thermal density matrix associated with either exterior.}. In this state the real-time bulk-to-bulk propagators,  satisfy the KMS relation
\cite{Son:2009vu,Hartle:1976tp,CaronHuot:2009iq}; in particular,
	\begin{equation}
	G_{\rm sym}^{bb}(\wn,\qn)
	=
	\frac{i}{2}
	\coth\!\left(\frac{\pi\wn}{2}\right)
	\left[
	G_R^{bb}(\wn,\qn)-G_A^{bb}(\wn,\qn)
	\right].
	\label{eq:symmetric-propagator}
\end{equation}
	The thermal factor in Eq.~\eqref{eq:symmetric-propagator} is the new state-dependent ingredient relative to the classical QNM problem. It sets the thermal weight of the QNM contributions in the loop and, at large loop momentum, gives rise to the Boltzmann suppression that is crucial for the late-time behavior discussed below. Together with the retarded bulk propagator $G_R^{bb}$ constructed in Appendix~\ref{App_bulk_sol}, Eq.~\eqref{eq:symmetric-propagator} specifies the real-time bulk propagators entering the loop.

	\subsection{The one-loop boundary correlator}
	\label{sec:one-loop-correlator}
	
	The boundary correlator can be obtained in two equivalent ways: by differentiating the bulk generating functional with respect to the boundary source,  which is referred to as the Gubser-Klebanov-Polyakov-Witten (GKPW) dictionary \cite{Witten:1998qj,Gubser:1998bc}, or by taking the appropriately rescaled and renormalized boundary limit of the bulk correlator \cite{Susskind:1998dq,Banks:1998dd}, which is referred to as the BDHM dictionary. The equivalence of these GKPW and BDHM dictionaries in interacting AdS field theory was established in Ref. \cite{Harlow:2011ke}. We use the latter representation below because it makes the projection of the bulk self-energy insertion onto the boundary correlator explicit. For the Lorentzian retarded implementation we use the standard
	real-time holographic prescriptions of
	Refs.~\cite{Son:2002sd,Skenderis:2008dh}.
	
	For the scalar field considered here,
 the ingoing bulk solution has the near-boundary expansion
	\begin{equation}
	u_{\rm in}(z;\mathfrak w,\kappa)
	=
	A(\mathfrak w,\kappa)\,z^{4-\Delta}
	+
	B(\mathfrak w,\kappa)\,z^\Delta+\cdots .
\end{equation}
where 	
$\Delta=2+\nu,$ and $\nu=\sqrt{4+m^2}=\sqrt5.$
	Here \(A\) is the non-normalizable source coefficient and \(B\) the normalizable response coefficient.
	Let \(\mathcal G_R^{bb}\) denote the full retarded bulk-to-bulk correlator and \(G_R^{bb}\) its tree-level part. We choose the normalization of the boundary operator such that 
	\begin{equation}\label{BDHM}
	\widehat G_R^\phi(\mathfrak w,\kappa)
	\equiv
	\mathcal A(2\nu)^2
	\operatorname*{Ren\,lim}_{z_1,z_2\to0}
	z_1^{-\Delta}z_2^{-\Delta}
	\mathcal G_R^{bb}
	(\mathfrak w,\kappa;z_1,z_2),
\end{equation}
	where the renormalized limit removes the non-normalizable and local contact contributions.
	
	At tree level, the renormalized boundary
	limit of the bulk propagator reproduces the standard holographic result (see App.~\ref{App_bulk_sol})
	\cite{Son:2002sd,Nunez:2003eq},
	\begin{equation}
	\widehat G_{R,0}^\phi(\mathfrak w,\kappa)
	=
	-2\nu\mathcal A\,
	\frac{B(\mathfrak w,\kappa)}
	{A(\mathfrak w,\kappa)}
	+\text{local terms}.
\end{equation}
	Note that  the boundary and bulk retarded Green functions therefore have poles at the same QNM frequencies, determined by $A(\wn_n(\kappa),\kappa)=0.$
	
	We now include the first bulk self-energy insertion. To first order in the retarded self-energy, the bulk propagator receives the correction \cite{Kamenev:2009jj}
	\begin{equation}\label{bulk_loop}
	\delta\mathcal G_R^{bb}
	(\mathfrak w,\kappa;z_1,z_2)
	=
	\int_0^1\frac{dz}{z^5}
	\int_0^1\frac{dz'}{z'^5}\,
	G_R^{bb}(\mathfrak w,\kappa;z_1,z)\,
	\Sigma_R(\mathfrak w,\kappa;z,z')\,
	G_R^{bb}(\mathfrak w,\kappa;z',z_2).
\end{equation}
         Taking one external bulk point to the boundary converts the corresponding retarded bulk-to-bulk propagator into the source-normalized retarded bulk-to-boundary propagator \(K_R^\phi\), as derived in Appendix~\ref{App_bulk_sol}. Applying this one-leg boundary limit to both external propagators in Eq.~\eqref{bulk_loop}, together with the definition \eqref{BDHM}, yields
	\begin{equation}\label{Dyson}
	\boxed{
		\delta\widehat G_R^\phi(\mathfrak w,\kappa)
		=
		\mathcal A
		\int_0^1\frac{dz}{z^5}
		\int_0^1\frac{dz'}{z'^5}\,
		K_R^\phi(z;\mathfrak w,\kappa)\,
		\Sigma_R(\mathfrak w,\kappa;z,z')\,
		K_R^\phi(z';\mathfrak w,\kappa).
	}
\end{equation}
	Equation \eqref{Dyson} is therefore the first retarded insertion in the bulk, projected onto the two boundary operators by the BDHM dictionary. It is represented schematically in Fig.~\ref{intro_loop}.
	
    	For the cubic interaction in Eq.~\eqref{action}, the retarded one-loop self-energy is built from two identical thermal bulk propagators. In the Schwinger–Keldysh \(r/a\) organization, the two assignments combine into one retarded and one statistical propagator \cite{Kamenev:2009jj,CaronHuot:2009iq}. Suppressing their arguments
	\begin{equation}\label{Sigma}
    \Sigma_R
    =
    \epsilon\,
    G_R^{bb}G_{\rm sym}^{bb}.
\end{equation}
    
    To display the frequency and momentum routing, we introduce the loop frequency \(\mathfrak w'\) and momentum \(\boldsymbol{\mathfrak q}\), and define
    $\mathfrak p\equiv
    \left|\boldsymbol\kappa-\boldsymbol{\mathfrak q}\right|.$
    Fourier transforming Eq.~\eqref{Sigma} gives
    	\begin{equation}	\label{eq:self-energy-fourier}
    \begin{aligned}
    	\Sigma_R(\mathfrak w,\kappa;z,z')
    	={}&
    	\epsilon
    	\int\frac{d^3\mathfrak q}{(2\pi)^3}
    	\int\frac{d\mathfrak w'}{2\pi}
         \,
    	G_R^{bb}(\mathfrak w',\mathfrak q;z,z')
    	G_{\rm sym}^{bb}(\mathfrak w-\mathfrak w',\mathfrak p;z,z').
    \end{aligned}
\end{equation}
   As we discuss in detail below, the convolution above is initially defined in the retarded analyticity domain, \(\operatorname{Im}\mathfrak w>0\), and is extended elsewhere by analytic continuation in the external frequency \cite{LeBellac:1996}.

    Short-distance singularities arise when the two bulk interaction vertices approach one another, producing the usual ultraviolet-local divergences of the one-loop correlator \cite{ChenLin:2018kmg,Abbasi:2022aao}. These divergences require regularization and local bulk counterterms. The corresponding terms renormalize local parameters and are analytic in the external frequency and momentum. At fixed renormalized parameters, they do not themselves modify the nonlocal QNM-pair singularities studied below; in particular, they cannot generate or cancel the discontinuity across a QNM-pair branch cut. This separation between ultraviolet-local contributions and the nonanalytic part of a loop correlator is analogous to that encountered in holographic long-time-tail calculations \cite{CaronHuot:2009iq}.

        With the one-loop representation established, we now turn to its nonanalytic dependence on the external frequency.

	\FloatBarrier
	\section{Signed QNM pairs and pinch singularities}
	\label{sec:signed-pairs}
	We now turn to the analytic structure generated by pairs of bulk QNM poles. At fixed loop momentum, pinching the internal-frequency contour produces a pole in the external frequency at a sum of two QNM frequencies. As the loop momentum varies, these poles trace QNM-pair frequency curves; the remaining momentum integration turns these loci into branch cuts and selects their branch points. Appendix~~\ref{sec:mix-general-coefficient} derives the coefficient with which each signed QNM pair contributes to the boundary correlator.
	
\subsection{From the loop integral to candidate branch points}
\label{sec:pair-geometry}

For fixed real loop momentum $\boldsymbol{\mathfrak q}$, consider the internal-frequency convolution in Eq.~\eqref{eq:self-energy-fourier}, inserted into the boundary correction \eqref{Dyson}. As long as the singularities of the integrand remain separated from the $\mathfrak w'$ contour, the contour can be smoothly deformed as the external frequency $\mathfrak w$ is analytically continued. An obstruction occurs when singularities approach the contour from opposite sides and no deformation can keep the contour between them. This is the standard pinch mechanism of analytic $S$-matrix theory 
\cite{Eden:1966asm,Landau:1959fi,Hannesdottir:2022xki}.  

The Landau approach \cite{Landau:1959fi} was introduced into relativistic hydrodynamic response functions in Ref.~\cite{Abbasi:2024hydro} \footnote{See Sec. IV B of Ref.~\cite{Abbasi:2024hydro}, especially Secs. IV B 1 and IV B 2, for the branch-cut construction and the Landau analysis, and App. C for a review of the standard relativistic Landau equations. Ref.~\cite{Abbasi:2024hydro} emphasizes that, unlike the usual relativistic on-shell condition \(q^2+m^2=0\), the hydrodynamic on-shell conditions used there are not Lorentz covariant.}, where it was adapted to non-Lorentz-invariant mode dispersion relations and used to determine branch-cut loci and threshold singularities directly from the hydrodynamic spectrum, without first evaluating the full loop integral (see also Ref.~\cite{Abbasi:2025teu})\footnote{More recently, Ref.~\cite{Wang:2026tzu} applied the same framework to nonlinear late-time relaxation for prescribed dispersion relations.}.
Here we use the same methodological idea in a different setting, where the internal excitations are gapped QNMs rather than hydrodynamic modes.

\begin{figure}[htbp]
	\centering\includegraphics[width=\textwidth]{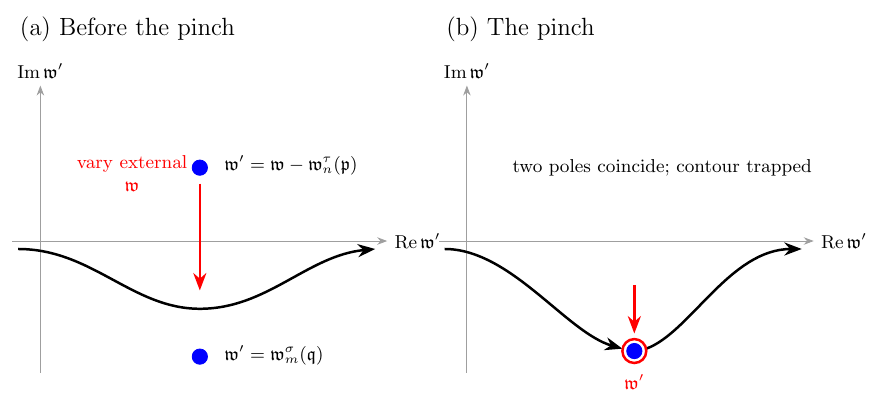}
\caption{
	Pinch mechanism at fixed loop momentum. The black curve denotes the deformed \(\mathfrak w'\) integration contour, and the blue dots denote the two QNM poles. As the external frequency \(\mathfrak w\) is varied, the shifted pole \(\mathfrak w-\mathfrak w_n^\tau(\mathfrak p)\) moves toward the pole \(\mathfrak w_m^\sigma(\mathfrak q)\) from the opposite side of the contour. When the two poles meet, the contour is trapped between them and becomes pinched. Here \(\mathfrak p=|\vextmom-\boldsymbol{\qn}|\).
}
	\label{fig:mix-frequency-pinch}
\end{figure}

\paragraph{Which poles can meet?}
For real \(\qn\equiv|\boldsymbol{\qn}|\), let \(\wn_m^+(\qn)\) denote a QNM pole of the retarded bulk propagator with positive real part. For real \(q\), reality of the bulk equations implies that the retarded QNM poles occur in reflected pairs, $\wn_m^-(\qn)=-[\wn_m^+(\qn)]^*$, where
\[
\qquad
\text{Re}\,\wn_m^+>0,\qquad
\text{Re}\,\wn_m^-<0,\qquad
\text{Im}\,\wn_m^\pm<0,
\qquad m=0,1,2,\ldots .
\]
The indices \(m\) and \(n\) label the QNM levels carried by the two internal propagators. Although both propagators belong to the same scalar field, their mode indices are independent. We choose \(\mathfrak w_m^\sigma(\mathfrak q)\) for the propagator carrying \((\mathfrak w',\mathfrak q)\) and \(\mathfrak w_n^\tau(\mathfrak p)\) for the one carrying \((\mathfrak w-\mathfrak w',\boldsymbol\kappa-\mathfrak q)\), where \(\sigma,\tau\in\{+,-\}\). In the \(\mathfrak w'\) plane, the corresponding pole locations are
\begin{equation}\label{Poles}
\wn'=\wn_m^\sigma(\qn),
\qquad
\wn'=\wn-\wn_n^\tau(\mathfrak p).
\end{equation}
The initial \(\wn'\) contour is the real-frequency contour in Eq.~\eqref{eq:self-energy-fourier}. Since we are interested in the retarded correlator, the external frequency is initially taken in the upper half-plane, \(\text{Im}\,\wn>0\), where causality guarantees analyticity. Thus, the first pole in Eq.\eqref{Poles} lies below the real \(\wn'\) axis, whereas the second pole lies above it.
We then analytically continue \(\wn\) away from the upper half-plane while deforming the \(\wn'\) contour continuously without allowing either pole to cross it. Each pole therefore remains on the same side of the deformed contour on which it started, as shown by panel (a), in Fig.~\ref{fig:mix-frequency-pinch}. A singularity can arise when the two poles meet and no contour can still be deformed between them.

The  singular contribution from the \(\wn'\) integral is then given by 
\begin{equation}\label{frequency_pinch_integral}
\left.
\int_{C_{\wn'}}\frac{d\wn'}{2\pi}\,
\frac{1}{
	[\wn'-\wn_m^\sigma(\qn)]
	[\wn-\wn'-\wn_n^\tau(\mathfrak p)]
}
\right|_{\rm sing}
=
-\,\frac{i}{
	\wn-\wn_m^\sigma(\qn)-\wn_n^\tau(\mathfrak p)
} .
\end{equation}
Here the sign follows from closing the contour in the lower half-plane, for which the orientation is clockwise.

\paragraph{From QNM-pair curves to branch points.}
This motivates defining the signed pair frequency
\begin{equation}
	\Omega_{mn}^{\sigma\tau}(\boldsymbol{\qn};\vextmom)
	=
	\wn_m^\sigma(|\boldsymbol{\qn}|)
	+
	\wn_n^\tau(|\vextmom-\boldsymbol{\qn}|), \qquad \sigma,\tau=\pm.
	\label{eq:general-pair-frequency}
\end{equation}
We henceforth set the external spatial momentum to zero, $\kappa=0$, for which
\begin{equation}\label{pole_collision}
	\Omega_{mn}^{\sigma\tau}(\qn)
=
\mathfrak w_m^\sigma(\qn)+\mathfrak w_n^\tau(\qn).
\end{equation}
As \(\qn\) varies continuously, this pole traces out a QNM-pair frequency curve. After the remaining \(\qn\) integration in Eq.~\eqref{eq:self-energy-fourier}, this continuous locus becomes the corresponding \textit{branch cut} of the boundary correlator, shown in Fig.~\ref{intro_overview}.   \textit{Branch points} occur where the inverse map $\qn(\mathfrak w)$ ceases to be locally single valued, in particular at endpoints or stationary points of the pair-frequency curve \cite{Landau:1959fi,Eden:1966asm}. Whether a candidate branch point is present in the boundary correlator also depends on the corresponding QNM residues, thermal weight, and radial projection. 



\paragraph{Generic points along the branch cut.} Suppose \(q_0>0\) and \(\Omega'(q_0)\neq0\). This corresponds to a generic point along the cut, away from an endpoint or stationary point. By the inverse-function theorem, \(\qn(\mathfrak w)\) is locally single-valued near \(\mathfrak w=\Omega(\qn_0)\).  The resulting momentum integral then takes the local form
\begin{equation}\label{I_Generic}
{\cal I}(\wn)
=
\int d\qn\,\frac{f(\qn)}{\wn-\Omega(\qn)},
\end{equation}
with $f$ smooth and including the radial measure, the two contour
continuations on opposite sides of the pole differ by the residue,
\begin{equation}
	\operatorname{Disc}{\cal I}(\Omega(\qn_0))
	=
	\pm 2\pi i\,
	\frac{f(\qn_0)}{\partial_{\qn}\Omega(\qn_0)}.
	\label{eq:regular-cut-discontinuity}
\end{equation}
The sign depends on which of the two analytic continuations across the cut is subtracted from the other and will be fixed below for the mixed imaginary-axis cut.

This jump does not imply square-root monodromy around every point of the
cut.  At a generic point along the cut, the momentum contour can be locally deformed and
each side continued analytically.  At a branch point, however, the inverse
map instead develops two local branches.  This distinction between a generic point of the cut and a genuine branch point will be important throughout the analysis below.

It should be noted that the thermal KMS factors also introduce Matsubara poles in the loop-frequency plane. These are included in the full contour continuation described in Appendix~\ref{app:fullSK}; the branch cuts discussed above arise specifically from pinches of QNM poles from the two internal propagators.

\subsection{The lowest mixed-sign branch cut}
\label{sec:mix-lowest}

	Applying the criterion of Sec.~\ref{sec:pair-geometry} to the \(00^{+-}\) channel, we find no finite real-$\qn$ stationary point over the computed range $0\le \qn\le1306$. We therefore focus on the endpoint $\qn=0$; the opposite limit $\qn\to\infty$, responsible for the accumulation at $\mathfrak w=0$, will be discussed later.

Using \(\wn_-(\qn)=-\wn_+(\qn)^*\), the lowest mixed-sign pair frequency is purely imaginary,
	\begin{equation}\label{mix_locus}
		\Omega(\qn)=\wn_+(\qn)+\wn_-(\qn)
		=2i\,\Im\wn_+(\qn).
	\end{equation}
	Because the QNM dispersion is even in $\qn$, the pair frequency near the
	endpoint $\wn_*=\Omega(0)$ has the expansion
	\begin{equation}\label{quadratic_expansion}
	\Omega(q)=\wn_*+a_{+-}\qn^2+O(\qn^4),
	\qquad
	a_{+-}=\frac12\Omega''(0)\neq0 .
	\end{equation}
	Hence,
	$q_\pm(\mathfrak w)\simeq
	\pm\sqrt{\frac{\mathfrak w-\mathfrak w_*}{a_{+-}}},$
	identifying \(\mathfrak w_*\) as the candidate square-root endpoint. Near this singular point, the monetum integral takes the followinmg form
		\begin{equation}\label{G_00_mix_1}
		G_{00}^{\rm mix}(\wn)
		=
		\epsilon
		\int_0^\infty
		\frac{\qn^2\,d\qn}{2\pi^2}
		\frac{H_{00}^{\rm mix}(\qn)}
		{\wn-\Omega(\qn)} .
	\end{equation}
	The whole information associatd with the pole resude, the KMS weight, and the radial vertex is encoded in $H_{00}^{\text{mix}}(\qn)$. See App.~\ref{sec:mix-general-coefficient}  for details.	
	
	To extract the singular behavior near the endpoint, let us take
	$\delta=\wn-\wn_* .$
	Only the small-$\qn$ region becomes singular as $\delta\to0$.  We may therefore
	restrict the integral to $0\le \qn\le\Lambda$, with $\Lambda$ fixed and small,
	and use $	H_{00}^{\rm mix}(\qn)=H_*+O(\qn^2),$ together with Eq.~\eqref{quadratic_expansion}.
	The singular part of Eq.~\eqref{G_00_mix_1} is then
	\begin{equation}
		\left.G_{00}^{\rm mix}(\wn)\right|_{\rm sing}
		=
		\frac{\epsilon H_*}{2\pi^2}
		\int_0^\Lambda
		\frac{\qn^2\,d\qn}{\delta-a_{+-}\qn^2}
		+\text{terms less singular near }\wn_* .
		\label{eq:local-threshold-integral}
	\end{equation}
		Defining
	$\eta=\sqrt{-\frac{\delta}{a_{+-}}},$
	the integral becomes $	-\frac{\epsilon H_*}{2\pi^2a_{+-}}
	\left[
	\Lambda-\eta\arctan\!\left(\frac{\Lambda}{\eta}\right)
	\right].$
	As $\eta\to0$, the term proportional to $\Lambda$ is analytic in
	$\wn$, whereas
	$\arctan(\Lambda/\eta)\to\pi/2$.  The leading nonanalytic contribution is
	therefore proportional to $\eta\propto\sqrt{\wn-\wn_*}$, giving
	\begin{equation}
		G_{00}^{\rm mix}(\wn)=G_{\rm hol}(\wn)
		+\mathcal C_*\sqrt{-\frac{\wn-\wn_*}{a_{+-}}}
		+O(\delta^{3/2}).
		\label{eq:mix-root}
	\end{equation}
with $\frac{\mathcal C_*}{\epsilon}=\frac{H_*}{4\pi a_{+-}}$. Here $G_{\rm hol}$ is analytic near $\wn_*$. Because of the square root, a nonzero $H_*$ establishes $\wn_*$ as a
branch point of the boundary correlator.

\paragraph{QNM-pair prediction.}
 Only at this stage do we insert the numerical QNM data. For the lowest mixed-sign pair,
	\begin{equation}
\mathfrak w_*=\mathfrak w_+(0)+\mathfrak w_-(0)
=-5.967880\,i .
	\end{equation}
The small-\(\qn\) dispersion gives 
	\begin{equation}
a_{+-}
=0.146432\,i\neq0 ,
	\end{equation}
confirming that the endpoint is quadratic. Evaluating the radial vertex, thermal factor, and QNM residues gives a finite nonzero coefficient,
	\begin{equation}
H_*\equiv H_{00}^{\rm mix}(0)\neq0 ,
	\end{equation}
with the individual ingredients given in App.~\ref{app:mixed_numerics}. Thus the coefficient $\mathcal C_*$ is nonzero and therefore, Eq.~\eqref{eq:mix-root} predicts a genuine branch point at $\wn_*$
with leading square-root behavior.


\paragraph{Blind test with the complete Schwinger–Keldysh loop.}
The analysis above isolates the local QNM-pair contribution and predicts both the endpoint \(\mathfrak w_*\) and its square-root character. We now test these predictions directly in the complete Schwinger–Keldysh loop, without replacing the internal propagators by their local QNM poles.

\begin{SCfigure}
	\centering
	\includegraphics[
	width=0.6\linewidth,
	trim={0 15bp 0 0},
	clip
	]{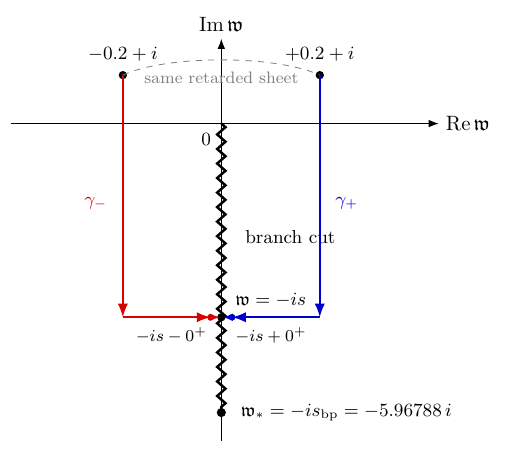}
	\vspace{-3mm}
	\caption{
		Two paths \(\gamma_-\) and \(\gamma_+\) in the external \(\mathfrak w\)-plane defining the two analytic continuations of the retarded correlator. Starting from points connected within the analytic upper half-plane, the paths approach the common point \(\mathfrak w=-is\) from opposite sides of the cut on the negative imaginary axis. Their difference defines the discontinuity measure $\Delta \widehat G_R^\phi(s)$ used in the blind test. The figure shows only the continuation paths in the external \(\mathfrak w\)-plane; for each path \(\gamma_\pm\), the associated transport of the loop-frequency contour in the \(\mathfrak w'\)-plane is described in App.~\ref{app:fullSK}.
	}
	\label{fig:gamma_paths}
\end{SCfigure}

We define the two lateral analytic continuations represented schematically by the paths \(\gamma_\pm\) in Fig.~\ref{fig:gamma_paths}. They start on the same retarded sheet and approach \(\mathfrak w=-is\) from opposite sides of the cut. Numerically, these continuations are implemented using a fixed reference contour for the internal frequency, with the appropriate complete-kernel residues added whenever a moving singularity crosses the reference contour; see App~\ref{app:fullSK}. We denote the difference of the two continuations by
	\begin{equation}
\Delta \widehat G_R^\phi(s)
\equiv
\widehat G_R^{\phi,\gamma_+}(-is)-\widehat G_R^{\phi,\gamma_-}(-is).
	\end{equation}
which measures the discontinuity across the cut.

\begin{figure}[t]
	\centering
	\includegraphics[width=0.8\linewidth,
	trim={0 1bp 0 0},
	clip
	]{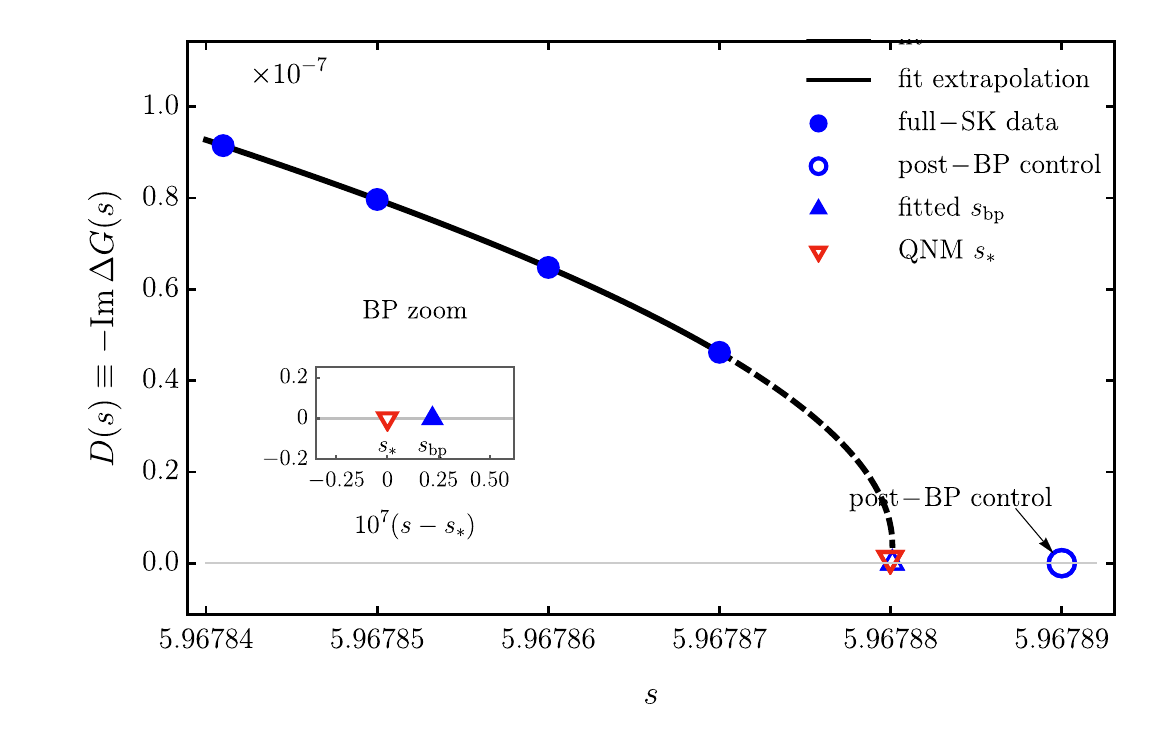}
	\vspace{-3mm}
	\caption{
		The blue points are complete SK data. The solid curve is a blind near-endpoint fit to a power-law singular contribution plus a smooth analytic background, given by Eq.~\eqref{fit} with \(s_{\rm bp}\) and \(\beta_{\text{bp}}\) free. The open point is a post-endpoint control constraining the smooth background and is not included in the fit. The inset compares the fitted branch point with the QNM prediction, which was not used in the fit.
	}
	\label{fig:blind-endpoint}
\end{figure}

Note that near the lowest \(00^{+-}\) endpoint, the complete discontinuity need not consist solely of the singular \(00\) contribution. As shown in Fig.~\ref{intro_overview}, higher QNM-pair cuts can pass through the same value of \(s\), but their contributions are analytic across the \(00\) endpoint and, as discussed below, are numerically negligible on the scale resolved here. We therefore write locally
	\begin{equation}\label{fit}
D(s)\equiv-\text{Im}\,\Delta \widehat G_R^\phi(s)
=
D_{\text{reg}}(s)+C_{\text{bp}}(s_{\rm bp}-s)^{\beta_{\text{bp}}},
\qquad s<s_{\rm bp},
	\end{equation}
where $D_{\text{reg}}(s)$ is analytic near $s_{\rm bp}$. For $s>s_{\rm bp}$, the $00$ singular contribution is absent while the smooth background remains \footnote{Strictly, this statement holds only for $s_{\rm bp}<s<s_{\text{next}}$, where $s_{\text{next}}>s_{\rm bp}$ denotes the next branch point on the negative imaginary axis. Since its location does not affect the late-time response discussed below, we do not determine $s_{\text{next}}$ here.}.
To determine the branch point without using the QNM prediction, we first scan \(D(s)\) over a broad interval and then refine the sampling around the endpoint. We fit the near-endpoint data to the right hand side expression in Eq.~\eqref{fit}. 
The fit gives (see Fig.~\ref{fig:blind-endpoint})
\begin{equation}
	\boxed{
		s_{\rm bp}=5.9678801\pm2\times10^{-7},
		\qquad
		\beta_{\text{bp}}=0.501\pm0.003 .
	}
	\label{eq:blind-result}
\end{equation}
This is in agreement with the independent QNM-pair prediction \(s_*=i\mathfrak w_*=5.967880\) and the square-root exponent $\beta_{\text{bp}}=1/2$.
We have therefore established the finite $00^{+-}$ branch point in two
independent ways: analytically from the QNM-pair singularity and numerically
from the complete Schwinger--Keldysh loop.  No additional finite branch point
is resolved in the numerical scan over $0.35\le s\le7.35$; the distinct
large-momentum accumulation at $s=0$ is analyzed below.

    The lowest mixed-sign cut studied above is one member of a much larger QNM-pair frequency network. More representative members are shown in Fig.~\ref{intro_overview}.

%
%
%

	\section{Accumulation of the mixed-sign tower at the origin}
	\label{sec:origin-accumulation}
	
	The lowest mixed-sign cut established in Sec.~\ref{sec:mix-lowest} is parametrized by the loop momentum \(\mathfrak q\in[0,\infty)\). Its finite endpoint at \(\mathfrak q=0\) was analyzed there; here we study the opposite limit, \(\mathfrak q\to\infty\). At large momentum, the two QNMs forming a mixed-sign pair approach opposite light-cone branches \cite{Festuccia:2008zx},
	\[
	\mathfrak w_m^+(\mathfrak q)\sim+\mathfrak q,
	\qquad
	\mathfrak w_n^-(\mathfrak q)\sim-\mathfrak q,
	\]
	so their leading terms cancel in the pair frequency. The remaining subleading contribution drives every  mixed-sign pair-frequency curve toward the origin of the complex \(\mathfrak w\) plane.
	
	The location of this accumulation alone does not determine its contribution to the real-time response. The KMS weight of the same large-\(\mathfrak q\) modes is exponentially suppressed, so the late-time behavior is controlled by the competition between the approach to \(\mathfrak w=0\) and the decreasing weight carried by the corresponding cuts. We first determine the large-\(\mathfrak q\) geometry of the mixed-sign QNM pairs and then their asymptotic weight. These two ingredients provide the input for the late-time analysis of Sec.~\ref{sec:real-time}.

	\subsection{Large-$\mathfrak q$ geometry of QNM pairs}
	
	The large-momentum asymptotics of black-brane QNMs were derived by
	Festuccia and Liu~\cite{Festuccia:2008zx}.  Specializing their result to the scalar
	model defined in Sec.~\ref{sec:linear-propagators}, for each fixed overtone index one finds
	\begin{equation}
		\mathfrak w_n^+(\mathfrak q)
		=
		\mathfrak q
		+
		\frac{\lambda_n}{2}e^{-i\pi/3}\mathfrak q^{-1/3}
		+
		O(\mathfrak q^{-5/3}),
		\qquad
		\lambda_n>0 .
		\label{eq:largeq-qnm}
	\end{equation}
	The constants $\lambda_n$ characterize the fixed-overtone asymptotics;
	the lowest value, needed below, is determined in App.~\ref{app:largeq-spectrum}.
	
	For a mixed-sign pair,
	\begin{equation}
		\Omega_{mn}^{+-}(\mathfrak q)
		=
		\mathfrak w_m^+(\mathfrak q)
		-
		\bigl[\mathfrak w_n^+(\mathfrak q)\bigr]^* .
	\end{equation}
	Substituting Eq.~\eqref{eq:largeq-qnm}, the leading light-cone terms
	cancel explicitly:
	\begin{align}
		\Omega_{mn}^{+-}(\mathfrak q)
		&=
		\left[
		\mathfrak q
		+
		\frac{\lambda_m}{2}e^{-i\pi/3}\mathfrak q^{-1/3}
		\right]
		-
		\left[
		\mathfrak q
		+
		\frac{\lambda_n}{2}e^{+i\pi/3}\mathfrak q^{-1/3}
		\right]
		+
		O(\mathfrak q^{-5/3})
		\nonumber\\
		&=
		d_{mn}\,\mathfrak q^{-1/3}
		+
		O(\mathfrak q^{-5/3}),
		\label{eq:largeq-pair}
	\end{align}
	where
		$d_{mn}
		\equiv
		\frac14
		\left[
		\lambda_m-\lambda_n
		-i\sqrt3\,(\lambda_m+\lambda_n)
		\right].$
	Hence, for every pair $(m,n)$,
	\begin{equation}
		\Omega_{mn}^{+-}(\mathfrak q)\longrightarrow0,
		\qquad
		\mathfrak q\longrightarrow\infty .
		\label{eq:origin-accumulation}
	\end{equation}
	
	The asymptotic direction is determined by $d_{mn}$.  Measuring the angle
	$\theta_{mn}$ from the negative imaginary axis,
	\begin{equation}
		\tan\theta_{mn}
		=
		\frac{\operatorname{Re}d_{mn}}
		{-\operatorname{Im}d_{mn}}
		=
		\frac{\lambda_m-\lambda_n}
		{\sqrt3\,(\lambda_m+\lambda_n)} .
		\label{eq:largeq-angle}
	\end{equation}
	Since $\lambda_m,\lambda_n>0$,
	\begin{equation}
			|\theta_{mn}|<\frac{\pi}{6}.
		\label{eq:wedge-bound}
	\end{equation}
	Thus every mixed-sign channel approaches the origin within a
	$60^\circ$ wedge centered on the negative imaginary axis, precisely the
	large-$\mathfrak q$ wedge indicated by the dashed cone in Fig.~\ref{intro_overview}.  This
	constraint applies only to the asymptotic region near $\mathfrak w=0$;
	finite-$\mathfrak q$ portions of the network need not lie inside the
	wedge.
	
	For the diagonal family $m=n$, the real part cancels identically and
	\begin{equation}
		\Omega_{nn}^{+-}(\mathfrak q)
		=
		-i\,a_n\,\mathfrak q^{-1/3}
		+
		O(\mathfrak q^{-5/3}),
		\qquad
		a_n\equiv\frac{\sqrt3}{2}\lambda_n .
		\label{eq:diagonal-largeq}
	\end{equation}
	The diagonal channels $00,11,22,\ldots$ therefore approach the origin
	along the negative imaginary axis.  For the lowest channel we write
	\begin{equation}
			a_\infty\equiv a_0=8.504951,
			\qquad
			\Omega_{00}^{+-}(\mathfrak q)
			=
			-i\,a_\infty\,\mathfrak q^{-1/3}
			+
			O(\mathfrak q^{-5/3}).
		\label{eq:lowest-largeq}
	\end{equation}
	Here the subscript $\infty$ indicates that $a_\infty$ governs the
	$\mathfrak q\to\infty$ asymptotics of the lowest channel; it is simply
	$a_0$, not an $n\to\infty$ limit.
	
	We next determine the asymptotic weight carried by these pair-frequency curves.

\subsection{Weight of the branch cut near the origin}
\label{sec:spectral-weight}
For the lowest mixed-sign cut, the weight entering the boundary correlator is encoded in \(H_{00}^{\rm mix}(\qn)\). Its large-\(\qn\) behavior follows separately from the thermal factor, the QNM residues, and the radial vertex.

Using Eqs.~\eqref{eq:largeq-qnm} and \eqref{eq:lowest-largeq}, the thermal factor
$T(\mathfrak q)
=
n_B\!\left(\mathfrak w_+(\mathfrak q)\right)
-
n_B\!\left(\mathfrak w_+(\mathfrak q)^*\right)$
has the asymptotic form
\begin{equation}
	T(\mathfrak q)
	\sim
	\frac{i\pi a_\infty}{2}\,
	\mathfrak q^{-1/3}e^{-\pi\mathfrak q}.
		\label{eq:T00-largeq}
\end{equation}
The exponential factor is the Boltzmann suppression of modes carrying
large internal momentum.

Combining Eq.~\eqref{eq:T00-largeq} with the large-$\mathfrak q$ boundary
layer of the QNM wavefunctions, $z\sim\mathfrak q^{-1/3}$ \cite{Festuccia:2008zx}, together with
the corresponding scaling of the QNM residues and radial vertex, we obtain
\begin{equation}\label{H_00_mix_box}
		H_{00}^{\rm mix}(\mathfrak q)
		\sim
		-i\,b\,
		\mathfrak q^{\,(2\nu-7)/3}
		e^{-\pi\mathfrak q},
		\qquad
		b=\frac{\pi a_\infty}{4}\mathcal R^2>0 .
\end{equation}
where \(\mathcal R=J_0/|N_0|\) is the positive constant defined in App.~\ref{app:largeq}.
Thus, although the lowest mixed-sign cut extends arbitrarily close to
$\mathfrak w=0$, the coefficient it carries becomes exponentially small
as $\mathfrak q\to\infty$.  

For every mixed-sign pair with nonzero leading projected coefficient, the same analysis gives
\begin{equation}\label{H_mn}
	H_{mn}^{\rm mix}(\mathfrak q)
	\sim
	B_{mn}\,
	\mathfrak q^{\,2(2\nu-7)/3}
	e^{-\pi\mathfrak q},
\end{equation}
with a channel-dependent coefficient $B_{mn}$.  The Boltzmann suppression
is therefore common to all fixed-index mixed-sign channels, while their
amplitudes depend on $(m,n)$.

     \subsection{Discontinuity near the origin}
     \label{sec:origin-disc}

     We now combine the large-$\mathfrak q$ trajectory of the lowest mixed-sign
     cut with its asymptotic weight.  At a regular point of the cut, its
     discontinuity follows from the momentum-pole residue derived in Sec.~\ref{sec:pair-geometry}, given by Eq.~\eqref{eq:regular-cut-discontinuity}
     \begin{equation}
     	\operatorname{Disc}G_{00}^{\rm mix}
     	\bigl(\Omega(\mathfrak q)\bigr)
     	=
     	-\frac{i\epsilon\,\mathfrak q^2}{\pi}\,
     	\frac{H_{00}^{\rm mix}(\mathfrak q)}
     	{\Omega'(\mathfrak q)} .
     	\label{Disc_H_00}
     \end{equation}
     On the other hand, near the origin, from Eq.~\eqref{eq:lowest-largeq},
with  $\mathfrak w=-is,$ ($s>0$),
     one has $ s\sim a_\infty\mathfrak q^{-1/3}.$
     Substitution into Eq.~\eqref{Disc_H_00} yields
     \begin{equation}
     		\operatorname{Disc}G_{00}^{\rm mix}(-is)
     		\sim
     		\frac{3i\epsilon}{4}\,
     		\mathcal R^2
     		\left(\frac{a_\infty}{s}\right)^{2\nu+3}
     		\exp\!\left[
     		-\frac{\pi a_\infty^3}{s^3}
     		\right].
     \end{equation}
     The cut therefore reaches arbitrarily close to $\mathfrak w=0$, while its
     discontinuity vanishes faster than any power of the distance from the
     origin.\textit{  In this sense it is exponentially flat}: for every  $M,$ $\lim_{s\to 0^+}s^{-M}\,
     \big|\operatorname{Disc}G_{00}^{\rm mix}(-is)\big|=0.$
     
     The same structure holds for every  mixed-sign pair.  Considering Eq.~\eqref{eq:largeq-pair}, and setting $ \rho\equiv|\mathfrak w|\to0$, 
     one has
     $\mathfrak q
     \sim
     \left(\frac{|d_{mn}|}{\rho}\right)^3.$
     Together with Eq.~\eqref{H_mn}, and wheever $B_{mn}\ne 0$, this gives, 
     \begin{equation}
     	\boxed{
     		\left|
     		\operatorname{Disc}G_{mn}^{\rm mix}
     		\right|
     		\propto
     		\rho^{-(2\nu+3)}
     		\exp\!\left[
     		-\frac{\pi|d_{mn}|^3}{\rho^3}
     		\right],
     		\qquad
     		\rho\to0 .
     	}
     \end{equation}
     Thus exponential flatness is a fixed-channel property of the mixed-sign
     accumulation.
     
     For the diagonal family, $d_{nn}=-ia_n$,
     and $a_0<a_1<a_2<\cdots$ (see App~\ref{app:largeq}).  Hence, for $n>0$,
     \begin{equation}
     	\frac{
     		\operatorname{Disc}G_{nn}^{\rm mix}(-is)}
     	{
     		\operatorname{Disc}G_{00}^{\rm mix}(-is)}
     	\propto
     	\exp\!\left[
     	-\frac{\pi(a_n^3-a_\infty^3)}{s^3}
     	\right]
     	\longrightarrow0,
     	\qquad
     	s\to0^+ .
     \end{equation}
     Among the diagonal mixed-sign channels, the $00$ cut is therefore
     asymptotically the least suppressed near the origin.

	\FloatBarrier
	\section{Real time response}
	\label{sec:real-time}
    
    Section~\ref{sec:origin-accumulation} determined the large-momentum geometry and asymptotic weight of the mixed-sign cuts. We now derive the corresponding late-time response. We first analyze the diagonal \(00^{+-}\) channel and then extend the saddle-point result to a general fixed mixed-sign pair. 

\subsection{Moving saddle and stretched-exponential relaxation}
\label{sec:moving-saddle}

\begin{figure}[t]
	\centering\includegraphics[width=\textwidth]{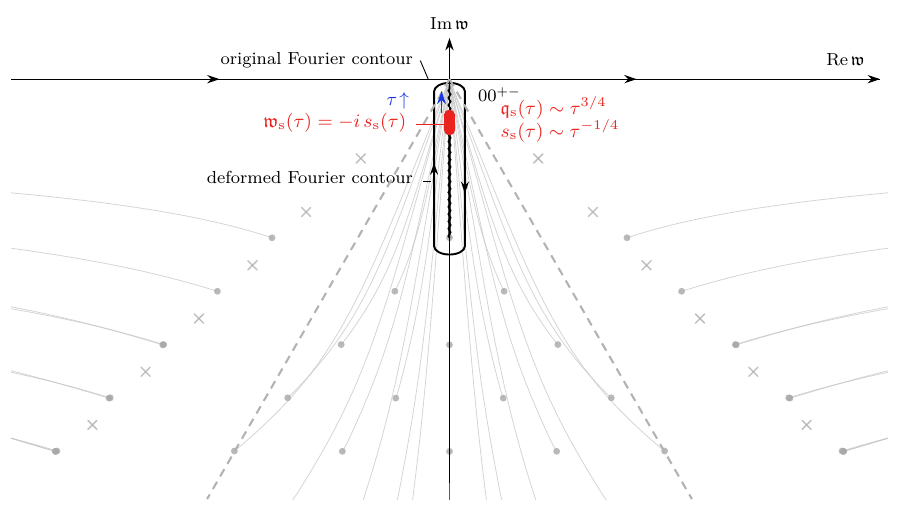}
	\caption{
		Late-time selection of the lowest mixed-sign cut in the complex-frequency plane. The QNM-pair frequency network is shown in light gray, while the \(00^{+-}\) cut is enclosed by the deformed inverse-Fourier contour. The exact contribution involves the entire cut, but at large \(\tau\) the integral is dominated by a narrow saddle region centered at \(\mathfrak w_s(\tau)=-is_s(\tau)\). As \(\tau\) increases, this region moves toward the origin, with \(\mathfrak q_s(\tau)\sim\tau^{3/4}\) and \(s_s(\tau)\sim\tau^{-1/4}\). The other structures are shown only for context and are not part of the mixed-sign saddle analysis.
	}
	\label{fig:late_time}
\end{figure}

We begin with the lowest mixed-sign channel. For \(\tau>0\),
\begin{equation}
	G_{00}^{\rm mix}(\tau)
	=
	-\frac{i\epsilon}{2\pi^2}
	\int_0^\infty d\qn\,
	\qn^2H_{00}^{\rm mix}(\qn)e^{-i\Omega(\qn)\tau}.
	\label{eq:G00-time}
\end{equation}
The full \(00^{+-}\) cut runs from the finite endpoint
\(\mathfrak w_*=\Omega(0)\) toward \(\mathfrak w=0\) as \(\qn\to\infty\). The question is which part of this cut controls the integral at late times. As shown below, the competition between the decreasing QNM-pair damping and the Boltzmann suppression selects a moving saddle at large $\qn$, so that the dominant region approaches $\mathfrak w=0$ as $\tau$ increases; see Fig.~\ref{fig:late_time}.

The same-sign pair frequencies also become weakly damped at large $\qn$, but their real parts grow as
$\Omega_{mn}^{++}(\qn)\sim 2\qn-ia_{mn}\qn^{-1/3},$
$\Omega_{mn}^{--}(\qn)\sim-2\qn-ia_{mn}\qn^{-1/3}.$
Their inverse transforms therefore contain rapidly varying phases \(e^{\mp2i\qn\tau}\) whose \(\qn\)-gradients remain nonzero at large $\qn$, so no moving large-$\qn$ saddle arises in the same-sign sector. We therefore restrict the present analysis to mixed-sign cuts, for which the leading light-cone terms cancel and \(\Omega_{mn}^{+-}(\qn)\to0\) as \(\qn\to\infty\).

Substituting Eqs.~\eqref{eq:lowest-largeq} and \eqref{H_00_mix_box} into Eq.~\eqref{eq:G00-time}, we obtain
\begin{equation}
G_{00}^{\rm mix}(\tau)
\sim
-\frac{\epsilon b}{2\pi^2}
\int_0^\infty d\mathfrak q\,
\mathfrak q^{p}
\exp\!\left[
-\pi\mathfrak q
-a_\infty\tau\mathfrak q^{-1/3}
\right],
\qquad
p\equiv\frac{2\nu-1}{3}.
	\label{eq:G00-saddle}
\end{equation}
The two terms in the exponent have opposite tendencies.  Increasing $\qn$
reduces the QNM damping, through $a_\infty\tau\mathfrak q^{-1/3}
$, but increases the
Boltzmann suppression $\pi \qn$.  Writing their sum as
$S(\mathfrak q;\tau)
\equiv
\pi\mathfrak q+a_\infty\tau\mathfrak q^{-1/3}$, the balance occurs at the saddle point
\begin{equation}
	\mathfrak q_{\rm s}(\tau)
	=
	\mathcal Q\,\tau^{3/4},
	\qquad
	\mathcal Q
	=
	\left(\frac{a_\infty}{3\pi}\right)^{3/4}.
	\label{eq:qs}
\end{equation}
Thus the momentum controlling the response moves to larger values as time
increases.  Equivalently, the corresponding point on the cut moves upward
toward $\wn=0$, as indicated in Fig.~\ref{fig:late_time}.
The late-time limit therefore drives the integral into the
same large-$\qn$ region in which Eqs.~\eqref{eq:lowest-largeq} and \eqref{H_00_mix_box} apply.

Evaluating the integral by steepest descent, we find for the \(00^{+-}\) channel
\begin{equation}
	\boxed{
G_{00}^{\rm mix}(\tau)
\sim
\epsilon A_{00}\tau^\alpha
e^{-C_{00}\tau^{3/4}},
\qquad
C_{00}
=
\frac{4\pi^{1/4}}{3^{3/4}}a_\infty^{3/4}>0,}
\end{equation}
with $\alpha=\frac{\sqrt5}{2}+\frac18.$
Thus a single fixed-index one-loop channel produces a stretched-exponential decay.

The same steepest-descent analysis applies to any fixed mixed-sign channel \((m,n)\) with nonzero leading amplitude. Using
$\kappa_{mn}\equiv i d_{mn}
=
\frac14\!\left[
\sqrt3(\lambda_m+\lambda_n)
+i(\lambda_m-\lambda_n)
\right],$
one obtains
\begin{equation}\label{G_mn_mix}
G_{mn}^{\rm mix}(\tau)
\sim
\epsilon A_{mn}\tau^\alpha e^{-C_{mn}\tau^{3/4}},
\qquad
C_{mn}
=
\frac{4\pi^{1/4}}{3^{3/4}}\kappa_{mn}^{3/4}.
\end{equation}

For the diagonal channels \(m=n\), \(C_{nn}\) is real and positive, so the stretched-exponential decay is nonoscillatory. For \(m\neq n\), the reflected channels \((m,n)\) and \((n,m)\) have complex-conjugate asymptotic contributions. Their sum therefore has an oscillatory stretched-exponential envelope,
\begin{equation}
G_{mn}^{\rm mix}+G_{nm}^{\rm mix}
\sim
2\epsilon |A_{mn}|\tau^\alpha
e^{-\operatorname{Re}C_{mn}\tau^{3/4}}
\cos\!\left(
\operatorname{Im}C_{mn}\tau^{3/4}-\varphi_{mn}
\right),
\end{equation}
where \(A_{mn}=|A_{mn}|e^{i\varphi_{mn}}\).

\subsection{Why the $00^{+-}$ channel is least suppressed}
	
	For a mixed-sign pair \((m,n)\), Eq.~\eqref{G_mn_mix} shows that the late-time suppression is controlled by \(\operatorname{Re}\mathcal C_{mn}\). Appendix \ref{app:fixed-channel-ordering} proves that
	\begin{equation}
		\operatorname{Re}\mathcal C_{mn}\geq \mathcal C_{00},
\end{equation}
	with equality only for \(m=n=0\). Consequently, every other mixed-sign channel with a nonzero leading amplitude is exponentially suppressed relative to the \(00^{+-}\) contribution as \(\tau\to\infty\).
	For the diagonal family,
	\begin{equation}
\kappa_{nn}=a_n=\frac{\sqrt3}{2}\lambda_n,
\end{equation}
	so that
	\begin{equation}
	\boxed{		
		\mathcal C_{nn}
		=		
		\mathcal C_{00}
		\left(
		\frac{\lambda_n}{\lambda_0}
		\right)^{3/4},
		\qquad		
		\mathcal C_{00}< \mathcal C_{11}< \mathcal C_{22}<\cdots .
	}
	\label{eq:C-diagonal}
\end{equation}
	Since \(\lambda_0<\lambda_1<\lambda_2<\cdots\), the \(00^{+-}\) channel is the least suppressed diagonal channel.
	
	\subsection{Late-time nonuniformity of the large-$N$ expansion}

	The accumulation at \(\mathfrak w=0\) has an important consequence for the late-time response. At classical order, the zero-momentum scalar correlator has a finite gap: its least-damped singularities are the lowest QNM poles, with damping rate	$\gamma_0\equiv-\operatorname{Im}\mathfrak w_0^+(0)>0.$
	At one loop, the mixed-sign QNM-pair cuts extend arbitrarily close to zero frequency, and the \(00^{+-}\) contribution is the least suppressed among all mixed-sign channels with nonzero leading weight. The one-loop analytic structure therefore no longer possesses the finite gap characteristic of the classical pole spectrum.
	
	In order to quantitatively compare the classical and one-loop contribustions at late-time, let us define the effective decay rate for an individual asymptotic contribution \(F(\tau)\)
	\begin{equation}
	\gamma_{\rm eff}[F](\tau)
	\equiv
	-\frac{d}{d\tau}\ln |F(\tau)|.
\end{equation}
	At classical order, the lowest QNM pole contributes
\begin{equation}
	G_{\rm QNM}^{(0)}(\tau)
	\sim
	Z_0e^{-i\mathfrak w_0^+(0)\tau},
	\qquad
	\gamma_0\equiv-\operatorname{Im}\mathfrak w_0^+(0)
	=2.983940\ldots ,
\end{equation}
and therefore
\begin{equation}
	\gamma_{\rm eff}\!\left[G_{\rm QNM}^{(0)}\right]=\gamma_0.
\end{equation}
	By contrast, Eq.~\eqref{G_mn_mix} gives
	\begin{equation}
		\gamma_{\rm eff}\!\left[G_{00}^{\rm mix}\right](\tau)
		=
		\frac{3\mathcal C_{00}}{4}\tau^{-1/4}
		-
		\left(\frac{\nu}{2}+\frac18\right)\tau^{-1}
		\longrightarrow0,
		\qquad \tau\to\infty .
\end{equation}
	The leading term is precisely the damping scale associated with the moving saddle.

	Thus the frequency scale controlling the response moves continuously toward zero frequency as \(\tau\) increases. At any nonzero loop-counting parameter \(\epsilon\), the ratio of the two decays behaves as
	\begin{equation}\label{instability}
\boxed{
	\frac{|G_{00}^{\rm mix}(\tau)|}
	{|G_{\rm QNM}^{(0)}(\tau)|}
	\propto
	\epsilon\,
	\tau^{\nu/2+1/8}
	\exp\!\left[
	\gamma_0\tau-\mathcal C_{00}\tau^{3/4}
	\right]
	\longrightarrow\infty .
}
\end{equation}

      The large-\(N\) expansion is therefore nonuniform at late times. At any fixed \(\tau\), the one-loop contribution is parametrically suppressed by \(\epsilon\sim1/N^2\). At fixed finite \(N\), however, its stretched-exponential time dependence decays more slowly than the classical QNM exponential and eventually becomes dominant.


	\subsection{Contrast with hydrodynamic long-time tails and the diffuson cascade}
	\label{Comparison_sec}
	
It is useful to distinguish the mechanism found here from the diffuson cascade of fluctuating hydrodynamics \footnote{We thank Richard Davison for discussions that motivated the comparison in Sec.~\ref{Comparison_sec} and the dimensional extension in Sec.~\ref{dimensional_sec}.}.
In fluctuating hydrodynamics, nonlinear mode coupling generates multiparticle branch cuts built from diffusive modes and the associated hydrodynamic long-time tails. At fixed nonzero external momentum \(k\), the simplest such contribution is the two-diffuson cut, whose threshold lies at
$\omega=-\frac{iDk^2}{2}$ \cite{ChenLin:2018kmg}.
Its real-time contribution contains the ordinary exponential \(e^{-Dk^2t/2}\), multiplied by a power-law prefactor. As \(k\to0\), the threshold reaches the origin and the exponential factor disappears, leaving the conventional power-law long-time tail \cite{Kovtun:2003vj}.

More generally, an \(n\)-diffuson contribution has threshold
$\omega_n=-\frac{iDk^2}{n},$
which moves progressively closer to the origin as \(n\) increases. Ref.~\cite{Delacretaz:2020nit} argued that, at sufficiently late times, the multiplicity that dominates the diffuson sum grows with time, producing the diffuson cascade and the associated stretched-exponential relaxation. For any fixed \(n\), however, the threshold remains at a finite distance from the origin when \(k\neq0\), and the corresponding contribution retains an ordinary exponential factor \(e^{-Dk^2t/n}\), multiplied by a power-law prefactor. The stretched exponential therefore arises only because the dominant multiplicity shifts to ever larger \(n\) as time increases. Moreover, the diffusion pole \(\omega=-iDk^2\) and all of these multiparticle thresholds approach \(\omega=0\) together as \(k\to0\), reflecting the gaplessness of the underlying conserved mode.

The mechanism identified here is qualitatively different. At vanishing external momentum, the classical scalar correlator is gapped: all of its QNM poles remain separated from the origin. Nevertheless, already at one-loop order, a single fixed-index mixed-sign QNM pair generates a branch cut extending to arbitrarily small frequency.

\textit{Thus the only singular structure reaching $\mathfrak w=0$ is generated by the one-loop cut; no tree-level pole approaches the origin.} Within this same one-loop channel, the competition between the decreasing QNM-pair damping and thermal Boltzmann suppression produces the moving saddle $q_s\sim\tau^{3/4}$ and the stretched-exponential decay
$G_{mn}^{\rm mix}(\tau)\sim
\tau^\alpha e^{-C_{mn}\tau^{3/4}}.$
Hence the effect found here is neither a conventional hydrodynamic long-time tail nor a diffuson cascade. It is the one-loop emergence of a continuum extending to zero frequency from a classically gapped nonhydrodynamic QNM spectrum. In this precise sense, the first finite-$N$ correction closes the gap set by  the classical QNM spectrum.

\subsection{Dimensional dependence of the stretched-exponential exponent}
	\label{dimensional_sec}
	The exponent $3/4$ in Eq.\eqref{instability} is not special to AdS\(_5\); it follows directly from the large-momentum approach of the QNMs to the light cone. For a planar AdS\(_{d+1}\) black brane, 
	Festuccia and Liu \cite{Festuccia:2008zx} found that the near-light-cone QNM damping in a \(d\)-dimensional CFT scales at large momentum as
	\begin{equation}
	\operatorname{Im}\omega_n\sim -q^{-\gamma_d},
	\qquad
	\gamma_d=\frac{d-2}{d+2},
    \end{equation}
	where \(d\) is the spacetime dimension of the boundary theory.
	
	For an opposite-sign QNM pair, the leading light-cone terms cancel, so that
	\begin{equation}
	\Omega_{mn}^{+-}(\mathfrak q)
	\sim
	d_{mn}^{(d)}\,\mathfrak q^{-\gamma_d}.
    \end{equation}
	Combining this with the large-\(\mathfrak q\) Boltzmann weight \(e^{-\pi\mathfrak q}\), the late-time exponent takes the form
	\begin{equation}
	\pi\mathfrak q
	+
	a_{mn}^{(d)}\tau\mathfrak q^{-\gamma_d}.
    \end{equation}
	The saddle therefore scales as
	\begin{equation}
	\mathfrak q_s(\tau)
	\sim
	\tau^{1/(1+\gamma_d)}
	=
	\tau^{(d+2)/(2d)},
    \end{equation}
	and hence
	\begin{equation}
	G_{mn}^{\rm mix}(\tau)
	\sim
	\tau^{\alpha_d}
	\exp\!\left[
	-C_{mn}^{(d)}
	\tau^{(d+2)/(2d)}
	\right].
    \end{equation}
	
	For the AdS\(_5\) case studied here, \(d=4\), this gives \(3/4\). In AdS\(_4\), \(d=3\), one instead finds
	\begin{equation}
	\gamma_3=\frac15,
	\qquad
	G^{\rm mix}(\tau)
	\sim
	\tau^{\alpha_3}e^{-C_3\tau^{5/6}}.
    \end{equation}
	Thus the stretched-exponential power is dimension dependent: it becomes closer to an ordinary exponential in lower dimension, while approaching \(1/2\) as \(d\to\infty\). This scaling argument fixes the stretched-exponential power; the algebraic prefactor and overall amplitude require a separate dimension-dependent analysis of the QNM residues and radial vertex.

	\section{Discussion and outlook}
	\label{sec:discussion}	
	At $N=\infty$, the retarded correlator considered here is meromorphic, and its late-time behavior is governed by isolated QNM poles. The first bulk-loop correction changes this analytic structure qualitatively: integrating QNM pairs over the continuous loop momentum generates branch cuts in addition to perturbatively dressing the classical poles. \textbf{The central result of this work is that the mixed-sign branch cuts extend to zero frequency, while the tree-level QNM poles themselves remain at finite nonzero frequencies. The first finite-\(N\) correction therefore closes the gap of the classical pole spectrum without requiring any classical QNM pole to approach zero frequency.} More specifically, we find square-root branch points at the $\qn=0$ endpoints of the QNM-pair cuts, an infinite family of mixed-sign cuts accumulating at $\mathfrak w=0$, and a stretched-exponential real-time contribution whose effective decay rate tends to zero. Among the mixed-sign channels, the $00^{+-}$ channel is asymptotically the least suppressed.
  

The accumulation of the mixed-sign cuts at \(\mathfrak w=0\) has a direct real-time consequence. In dimensionless time \(\tau=\pi Tt\), the leading classical contribution and the one-loop \(00^{+-}\) contribution behave as
\[
G_{\text{QNM}}^{(0)}(\tau)
\sim
e^{-\gamma_0\tau},
\qquad
 G^{\text{mix}}_{00}(\tau)
\sim
\frac{1}{N^2}\,
\tau^{\nu/2+1/8}
e^{-\mathcal C_{00}\tau^{3/4}}.
\]
At every fixed \(\tau\), the one-loop contribution is parametrically suppressed by \(1/N^2\). At fixed finite \(N\), however, its stretched-exponential decay is slower than the classical QNM exponential, so it eventually dominates beyond a crossover time \(\tau_{\rm cross}\) \footnote{A related subexponential relaxation mechanism was recently found in classical AdS dynamics, where the decreasing damping of large-momentum QNMs competes with the exponentially suppressed Fourier spectrum of real-analytic initial data \cite{Crump:2026wir}. In the present problem, the exponential weight instead originates intrinsically from the thermal KMS factor of the one-loop correlator, and the weakly damped object is a mixed-sign QNM pair.}. The large-\(N\) expansion is therefore nonuniform at late times. This does not imply that the one-loop result gives the ultimate finite-\(N\) asymptotic response: higher-loop multiparticle structures and the sum over the QNM tower may eventually become important. This nonuniformity is reminiscent of the broader distinction emphasized by Witten between finite-\(N\) black-hole physics and a naive strict large-\(N\) limiting description~\cite{Witten:2021unn,Schlenker:2022dyo}, although the mechanism considered there is different from the dynamical effect found here \footnote{We thank Amos Yarom for bringing the idea of Ref.~\cite{Witten:2021unn} to our attention.}. 

Accordingly, the stretched exponential should be viewed as an intermediate one-loop late-time regime. Its upper extent is limited independently by the breakdown of the $1/N$ expansion and by the range of validity of the two-derivative bulk theory. Since the dominant momentum grows as $q_s(\tau)\sim\tau^{3/4}$, the moving saddle eventually probes momentum scales at which the two-derivative bulk description need no longer be reliable \footnote{We thank Sašo Grozdanov for raising the question of ultraviolet sensitivity.}. Denoting these two limiting times by $\tau_{\rm higher,loop}$ and $\tau_{\rm UV}$, respectively, a parametric window for the one-loop stretched-exponential regime exists when
\[
\tau_{\rm cross}\lesssim\tau\ll
\min\!\left(\tau_{\rm UV},\tau_{\rm higher\ loop}\right).
\]
Beyond this window, higher orders in $1/N$ or ultraviolet modifications of the bulk theory may alter the response. Determining the ultimate finite-$N$ asymptotic behavior therefore lies beyond the present one-loop calculation.

A related mechanism for subexponential relaxation was recently found in
classical AdS dynamics, where the decreasing damping of large-momentum QNMs
competes with the exponentially suppressed Fourier spectrum of real-analytic
initial data~\cite{Crump:2026wir}.  The origin of the exponential weight in
the present problem is different: it is fixed intrinsically by the thermal
KMS factor of the one-loop correlator.  Moreover, the weakly damped object
here is a mixed-sign QNM pair.  Its large-momentum accumulation at
$\wn=0$ first generates an exponentially flat branch-cut discontinuity,
whose inverse Fourier transform then produces the stretched-exponential
late-time response.
	
	
	The point accumulation at \(\mathfrak w=0\) is special to vanishing external spatial momentum. At finite \(\boldsymbol{\mathfrak \kappa}\equiv\frac{\boldsymbol{k}}{\pi T},\)
	for $\mathfrak q\gg\mathfrak \kappa$, 
	\[
	\Omega_{mn}^{+-}
	\sim
	\mathfrak \kappa\cos\theta
	+
	d_{mn}\mathfrak q^{-1/3},
	\]
	where $\theta$ denote the angle between the loop momentum $\boldsymbol{\qn}$ and the external momentum $\boldsymbol\kappa$. 
	Thus, the \(\kappa=0\) accumulation point broadens asymptotically into the interval
	\[
	-\mathfrak \kappa\leq\operatorname{Re}\mathfrak w\leq\mathfrak \kappa.
	\]
	This is only a kinematic accumulation set. Determining the physical singularities at finite \(\mathfrak \kappa\) requires the full angular momentum integral, an analysis of the corresponding momentum-contour pinches, and boundary projection. Whether the late-time saddle retains the same \(\tau^{3/4}\) structure is likewise an open question.
	
	An important open problem is the fate of the QNM-pair accumulation beyond one loop. Higher-loop diagrams generate multiparticle QNM cuts, and their combined contribution may modify or require a resummation of the one-loop stretched-exponential regime.  A complementary extension is to conserved-current and stress-tensor correlators. In those channels, hydrodynamic and nonhydrodynamic modes coexist, allowing the mechanism found here to be compared directly with conventional hydrodynamic long-time tails. This would clarify how gapless small-momentum fluctuations and the large-momentum QNM-pair accumulation coexist and compete within the same correlator.

	It would also be interesting to investigate whether and how QNM-pair cuts appear in two-sided thermal correlators. At the classical level, such correlators can be written in terms of the QNM pole spectrum \cite{Dodelson:2023vrw}. If finite-$N$ branch cuts are also present, the pole spectrum alone may no longer be sufficient, and the discontinuities across the cuts may provide additional spectral data. This could clarify how the classical QNM product formula is modified by finite-$N$ effects.
	
	Together, these directions could help assess how robust the classical QNM description of thermal relaxation remains once finite-$N$, finite-coupling, and hydrodynamic-fluctuation effects are included.

     \section*{Acknowledgments}
     We are grateful to Sašo Grozdanov, Richard Davison, and Dirk Rischke for helpful comments on the manuscript.  We acknowledge the use of ChatGPT (OpenAI) for language editing, and assistance with implementing and debugging part of the Mathematica code used in this work. N.A. acknowledges support from the National Natural Science Foundation of China under Grant No. 12575142, and from the “Double First-Class” start-up funding of Lanzhou University, China, under Grant No. 561119208.
     \appendix
	
	\section{Bulk solutiosn and QNMs}
	\label{App_bulk_sol}
	With Fourier dependence
	$e^{-i\mathfrak w \tau+i\boldsymbol{\kappa}\cdot\mathbf x}$
	and $\kappa=|\boldsymbol{\kappa}|$, the radial Klein--Gordon equation is
	$\mathcal L_{\mathfrak \wn,\kappa}\big[u(\wn,\qn; z)\big]=0$
	with 
	\begin{equation}
		\mathcal L_{\mathfrak \wn,\kappa}=\partial_z\!\left(\frac{f}{z^3}\partial_z\right)
		+
		\frac{\mathfrak w^2}{z^3f}
		-
		\frac{\kappa^2}{z^3}
		-
		\frac{m^2}{z^5}\,.
		\label{eq:radial-operator}
	\end{equation}
	Near the boundary, inserting $u\sim z^\lambda$ gives
	\begin{equation}
		\Delta(\Delta-4)=1,
		\qquad
		\Delta=2+\sqrt{5},
		\qquad
		\nu\equiv\Delta-2=\sqrt5 ,
		\label{eq:dimension}
	\end{equation}
	as in the standard AdS/CFT scalar dictionary
	\cite{Gubser:1998bc,Witten:1998qj}.
	We choose two boundary-normalized homogeneous solutions \cite{Skenderis:2002wp,CasalderreySolana:2011us}
	\begin{equation}
		u_S(z)=z^{4-\Delta}(1+\cdots),
		\qquad
		u_N(z)=z^\Delta(1+\cdots),
		\label{eq:boundary-solutions}
	\end{equation}
	where $u_S$ carries the source falloff and $u_N$ is normalizable.

	At the horizon, the retarded and advanced boundary conditions select,
	respectively \cite{Son:2002sd,Herzog:2002pc},
	\begin{equation}
		u_{\rm in}\sim(1-z)^{-i\wn/4},
		\qquad
		u_{\rm out}\sim(1-z)^{+i\wn/4}.
		\label{eq:horizon-solutions}
	\end{equation}
	Continuing the ingoing solution to the boundary gives
	\begin{equation}
		u_{\rm in}(\wn,\kappa;z)
		=
		A(\wn,\kappa)\,u_S(z)
		+
		B(\wn,\kappa)\,u_N(z).
		\label{eq:source-response}
	\end{equation}
	The coefficient $A$ is the boundary-source coefficient and $B$ the normalizable response.
	
	We next construct the retarded bulk-to-bulk Green function using the
	standard normalizable--ingoing Wronskian representation
	\cite{Son:2009vu}.  It obeys
	\begin{equation}
		{\cal L}_{\wn, \qn}
		G_R^{bb}(\wn,\kappa;z,z')
		=
		\delta(z-z'),
		\label{eq:bulk-green-eq}
	\end{equation}
	together with a normalizable boundary condition at $z=0$ and an ingoing
	condition at the horizon.  For $z\neq z'$ the right-hand side vanishes, so
	the Green function must solve the homogeneous radial equation on either side
	of $z'$.  The endpoint conditions therefore require $u_N$ on the side
	containing the AdS boundary and $u_{\rm in}$ on the side containing the
	horizon.  Continuity at $z=z'$ then fixes the form
	\begin{equation}
		G_R^{bb}(\wn,\kappa;z,z')
		=
		C(\wn,\kappa)\,
		u_N(z_<)\,u_{\rm in}(z_>),
		\label{eq:bulk-green-form}
	\end{equation}
	where $	z_<\equiv\min(z,z'),$ and 
	$z_>\equiv\max(z,z').$ The normalization follows from the delta function.  Integrating
	Eq.~\eqref{eq:bulk-green-eq} across an infinitesimal interval around $z'$ gives
	the derivative jump
	\begin{equation}
		\left[
		\frac{f}{z^3}\partial_zG_R^{bb}
		\right]_{z=z'-0}^{z=z'+0}
		=1.
		\label{eq:jump-condition}
	\end{equation}
	Substitution of Eq.~\eqref{eq:bulk-green-form} yields
	\begin{equation}
		C(\wn,\kappa)\,W_R(\wn,\kappa)=1,
		\qquad
		W_R
		\equiv
		\frac{f}{z^3}
		\left(
		u_N\partial_zu_{\rm in}
		-
		u_{\rm in}\partial_zu_N
		\right).
		\label{eq:retarded-wronskian}
	\end{equation}
	Since $u_N$ and $u_{\rm in}$ solve the same homogeneous equation, it is easy to show that 
	$\partial_zW_R=0$, so the Wronskian is independent of $z$.  Hence
	\begin{equation}
		G_R^{bb}(\wn,\kappa;z,z')
		=
		\frac{u_N(z_<)\,u_{\rm in}(z_>)}{W_R(\wn,\kappa)}.
		\label{eq:bulk-retarded-propagator}
	\end{equation}
	The relation to QNMs follows immediately. Substituting Eq.~\eqref{eq:source-response} into the Wronskian, the term proportional to $B$ vanishes because $W[u_N,u_N]=0$, while the source term gives
	\begin{equation}
		W_R(\wn,\kappa)
		=
		A(\wn,\kappa)\,W[u_N,u_S]
		=
		-2\nu\,A(\wn,\kappa).
		\label{eq:wr-source}
	\end{equation}
	The poles of the retarded bulk Green function occur when the Wronskian in Eq. (A.10) vanishes. Equation~\eqref{eq:wr-source} shows that this happens precisely when
	$A(\mathfrak w,\kappa)=0.$
	At such a frequency, the ingoing solution contains no non-normalizable component and is therefore also normalizable at the AdS boundary. This is precisely the QNM condition. We thus denote its solutions by
	\begin{equation}		\label{eq:qnm-condition}
	A\!\left(\mathfrak w_n(\kappa),\kappa\right)=0.
	\end{equation}
		\textit{A QNM frequency $\mathfrak w_n(\kappa)$ is defined by the existence of a nontrivial solution that is simultaneously ingoing at the horizon and normalizable at the boundary.} 
	At a QNM frequency, the ingoing and normalizable solutions become linearly dependent,
	\begin{equation}		\label{eq:qnm-profile}
	u_{\rm in}\!\left(\mathfrak w_n(\kappa),\kappa;z\right)
	=
	B_n(\kappa)\,
	u_N\!\left(\mathfrak w_n(\kappa),\kappa;z\right),
	\qquad
	B_n(\kappa)\equiv
	B\!\left(\mathfrak w_n(\kappa),\kappa\right).
	\end{equation}
	Hence the Wronskian vanishes and the retarded bulk-to-bulk propagator develops a pole. For a simple QNM zero,
	\begin{equation}
	A(\mathfrak w,\kappa)
	=
	\bigl[\mathfrak w-\mathfrak w_n(\kappa)\bigr]
	\left.
	\partial_{\mathfrak w}A(\mathfrak w,\kappa)
	\right|_{\mathfrak w=\mathfrak w_n(\kappa)}
	+
	O\!\left(
	[\mathfrak w-\mathfrak w_n(\kappa)]^2
	\right).
	\end{equation}
	and the full bulk retarded propagator admits the local Laurent expansion
	\begin{equation}
		G_R^{bb}(\wn,\kappa;z,z')
		=
		\frac{r_n(\kappa;z,z')}{\wn-\wn_n(\kappa)}
		+
		G_{\rm reg}^{(n)}(\mathfrak w,\kappa;z,z'),
		\label{eq:bulk-qnm-pole}
	\end{equation}
	where \(G_{\rm reg}^{(n)}\) is analytic in a neighborhood of \(\mathfrak w_n\).
	The residue reads
	\begin{equation}
		r_n(\kappa;z,z')
		=
		\frac{u_n(\kappa,z)\,u_n(\kappa,z')}{D_n(\kappa)},
		\qquad
		D_n(\kappa)
		=
		-2\nu B_n(\kappa)
		\,\partial_\wn A(\wn_n(\kappa),\kappa),
		\label{eq:qnm-residue}
	\end{equation}
	where $u_n\equiv u_{\rm in}(\wn_n)=B_nu_N(\wn_n)$. We will use this local pole form below in the neighborhood of the frequency pinches.
	
	The advanced propagator is obtained by replacing the ingoing solution by the
	outgoing one,
	\begin{equation}
		G_A^{bb}(\wn,\kappa;z,z')
		=
		\frac{u_N(z_<)\,u_{\rm out}(z_>)}{W_A(\wn,\kappa)},
		\label{eq:bulk-advanced-propagator}
	\end{equation}
	and for real $\wn$ and $\kappa$,
	\begin{equation}
		G_A^{bb}(\wn,\kappa)
		=
		\bigl[G_R^{bb}(\wn,\kappa)\bigr]^*.
		\label{eq:ra-conjugation}
	\end{equation}
	Away from the real axis the two are understood as their respective
	causal analytic continuations \cite{Herzog:2002pc,Skenderis:2008dh}.
	
	    For fixed \(z>0\), we take the second radial point of the bulk-to-bulk propagator to the AdS boundary, \(z'\to0\). Then \(z_< = z'\) and \(z_> = z\) in Eq.~\eqref{eq:bulk-retarded-propagator}. Using $u_N(z')\sim z'^\Delta$ and \(W_R=-2\nu A\) then gives
	\begin{equation}\label{limi_G_R_bb}
		\lim_{z_1\to0}
		z_1^{-\Delta}
		G_R^{bb}(\mathfrak w,\kappa;z_1,z)
		=
		-\frac{1}{2\nu}\,
		K_R^\phi(z;\mathfrak w,\kappa),
	\end{equation}
	where
	\begin{equation}\label{K_phi_R}
		K_R^\phi(z;\mathfrak w,\kappa)
		\equiv
		\frac{u_{\rm in}(z;\mathfrak w,\kappa)}
		{A(\mathfrak w,\kappa)}
	\end{equation}
	is the retarded bulk-to-boundary propagator normalized to unit boundary source \footnote{
		Unlike the bulk-to-bulk Green function, which is the inverse of the radial wave operator and therefore has a delta-function source at \(z=z'\), the bulk-to-boundary propagator solves the homogeneous radial equation. The retarded condition selects the ingoing solution \(u_{\rm in}\), while division by its non-normalizable coefficient \(A(\mathfrak w,\kappa)\) normalizes the boundary source to unity, giving Eq.~\eqref{K_phi_R}. By contrast, the bulk-to-bulk Green function must also satisfy the normalizable AdS boundary condition on the side of the source point, which is why the solution $u_N$ enters its construction.}. 
	
	\section{QNM-pair coefficients}
	\label{App_B}

		\FloatBarrier
	\subsection{Coefficient of a general signed-pair singularity}
	\label{sec:mix-general-coefficient}
	The pair-frequency geometry determines where a candidate cut can occur, but not whether it survives in the boundary correlator. Its coefficient is controlled by three ingredients: the QNM pole residues of the two internal propagators, the thermal KMS weight, and the radial vertex coupling the pair to the external boundary operator. In this subsection we set $\kappa=0$, while keeping $m,n,\sigma,\tau$ general.

	\paragraph{First ingredient: pole residues.}
	Restoring the branch label in Eq.~\eqref{eq:qnm-residue}, we define 
	\begin{equation}
		u_n^\tau(\qn,z)\equiv u_{\rm in}(z;\wn_n^\tau(\qn),\qn),
		\qquad
		B_n^\tau(\qn)\equiv B(\wn_n^\tau(\qn),\qn).
	\end{equation}
	and 
	\begin{equation}
		D_n^\tau(q)
		=
		-2\nu B_n^\tau(q)\,
		\partial_{\mathfrak w}A(\mathfrak w_n^\tau(q),q).
	\end{equation}
	The Laurent residue of the bulk propagator at the selected QNM pole is therefore
	\begin{equation}
		r_n^\tau(q;z,z')
		=
		\frac{u_n^\tau(q,z)u_n^\tau(q,z')}{D_n^\tau(q)}.
		\label{eq:mix-pole-normalization}
	\end{equation}
	For real \(\mathfrak q\), the retarded propagator obeys
	\begin{equation}\label{G_R_bb_symmetry}
		G_R^{bb}(-\wn^*,\qn;z,z')
		=
		\bigl[G_R^{bb}(\wn,\qn;z,z')\bigr]^* .
	\end{equation}
	Consequently,
	\begin{equation}
		\mathfrak w_n^-
		=
		-\mathfrak w_n^{+*},
		\qquad
		u_n^-
		=
		u_n^{+*},
		\qquad
		D_n^-
		=
		-D_n^{+*},
		\qquad
		r_n^-
		=
		-r_n^{+*}.
		\label{eq:residue-reality}
	\end{equation}
	These relations will be useful when the two reflected QNM branches are
	combined below.
	
	\paragraph{Second ingredient: the thermal KMS weight.}
	The statistical propagator is related to the retarded and advanced
	propagators by Eq.~\eqref{eq:symmetric-propagator}.
	For brevity, let us define
	$c(\wn)\equiv \coth\!\left(\frac{\pi\wn}{2}\right).$
	Using the KMS relation in the retarded self-energy, the part containing two retarded propagators is
	\begin{equation}
		\left.
		\frac{\Sigma_R}{\epsilon}
		\right|_{RR}
		=
		\frac{i}{4}
		\left[
		c(\mathfrak w')
		+
		c(\mathfrak w-\mathfrak w')
		\right]
		G_R^{bb}(\mathfrak w',\mathfrak q)\,
		G_R^{bb}(\mathfrak w-\mathfrak w',\mathfrak q).
		\label{eq:RR-thermal}
	\end{equation}
	The factor \(1/4\) includes the identical-line symmetry factor \(1/2\) of the \(\phi^3\) bubble. The mixed retarded–advanced terms do not generate the QNM-pair pinch considered in Sec.~\ref{sec:pair-geometry}, because their relevant poles remain on the same side of the internal-frequency contour.
	
	Assuming that the thermal factors are regular at the selected QNM frequencies,  Eq.~\eqref{eq:general-pair-frequency} gives the
	pair weight
	\begin{equation}
		T_{mn}^{\sigma\tau}(\qn,p)
		=
		\frac14
		\left[
		c\!\left(\wn_m^\sigma(\qn)\right)
		+
		c\!\left(\wn_n^\tau(\mathfrak p)\right)
		\right].
		\label{eq:pair-thermal-coth}
	\end{equation}
	Using
	$c(\wn)=1+2n_B(\wn),$
	and 
	$n_B(\wn)=\frac{1}{e^{\pi\wn}-1},$
	this becomes
	\begin{equation}
		T_{mn}^{\sigma\tau}(\qn,\mathfrak p)
		=
		\frac12
		\left[1+n_B\!\left(\wn_m^\sigma(\qn)\right)
		+n_B\!\left(\wn_n^\tau(\mathfrak p)\right)\right].
		\label{eq:pair-thermal}
	\end{equation}
	For a mixed-sign channel 
	\begin{equation}
		T_{mn}^{+-}(\qn)
		=
		\frac12
		\left[n_B\!\left(\wn_m^+(\qn)\right)
		-
		n_B\!\left(\wn_n^+(\qn)^*\right)\right].
		\label{eq:mixed-thermal}
	\end{equation}
	Note that for \(m=n\), this quantity is purely imaginary and is generically nonzero.
	
	\paragraph{Third ingredient: the radial vertex.}
	After inserting the factorized QNM residues, the two radial integrations
	in the boundary correlator separate:
	\begin{align}\label{factorized_correlator}
		&\int_0^1\frac{dz}{z^5}\int_0^1\frac{dz'}{z'^5}\,
		K_R^\phi(z;\wn,0)K_R^\phi(z';\wn,0)\,
		r_m^\sigma(\qn;z,z')r_n^\tau(\qn;z,z')
		\nonumber\\
		&\qquad =
		\frac{1}{D_m^\sigma(\qn)D_n^\tau(\qn)}
		\left[
		\int_0^1\frac{dz}{z^5}\,
		K_R^\phi(z;\wn,0)
		u_m^\sigma(\qn,z)u_n^\tau(\qn,z)
		\right]
		\left[
		\int_0^1\frac{dz'}{z'^5}\,
		K_R^\phi(z';\wn,0)
		u_m^\sigma(\qn,z')u_n^\tau(\qn,z')
		\right].
	\end{align}
	The two factors are identical, so we define the ``radial vertex''
	\begin{equation}
		v_{mn}^{\sigma\tau}(\wn,\qn)
		=
		\int_0^1\frac{dz}{z^5}\,
		K_R^\phi(z;\wn,0)\,
		u_m^\sigma(\qn,z)u_n^\tau(\qn,z).
		\label{radial_overlap}
	\end{equation}
	The integral in Eq.~\eqref{radial_overlap} converges at the AdS boundary. At the QNM frequencies, however, it need not converge at the horizon when evaluated directly. We therefore first define the radial vertex in a frequency domain where the horizon integral converges and then analytically continue it to the QNM frequencies, as detailed in Appendix~\ref{app:mixed_numerics}.

	Using Eq.~\eqref{radial_overlap}, Eq.~\eqref{factorized_correlator} becomes
	$\bigl[v_{mn}^{\sigma\tau}(\wn,\qn)\bigr]^2/
	[D_m^\sigma(\qn)D_n^\tau(\qn)]$.
	Combining the three ingredients gives
	\begin{equation}
		\left.G_{mn}^{\sigma\tau}(\wn)\right|_{\rm sing}
		=
		\epsilon\int_0^\infty\frac{\qn^2\,d\qn}{2\pi^2}\,
		\frac{H_{mn}^{\sigma\tau}(\qn)}
		{\wn-\Omega_{mn}^{\sigma\tau}(\qn)}.
		\label{eq:general-pair-singular}
	\end{equation}
	where 
	\begin{equation}
		H_{mn}^{\sigma\tau}(\qn)
		=
		\frac{
			T_{mn}^{\sigma\tau}(\qn,\qn)
			\left[
			v_{mn}^{\sigma\tau}
			\bigl(\Omega_{mn}^{\sigma\tau}(\qn),\qn\bigr)
			\right]^2
		}{
			D_m^\sigma(\qn)D_n^\tau(\qn)
		}.
		\label{eq:H-general}
	\end{equation}
	Note that the square is not an absolute square, since both external legs are
	retarded.

	Equations~\eqref{eq:general-pair-singular} and \eqref{eq:H-general} reduce the
	problem to three quantities: the QNM residue normalizations \(D_m^\sigma\) and \(D_n^\tau\), the thermal KMS weight,
	and the radial vertex with the external boundary operator. The next subsection specializes this result to the lowest mixed-sign channel.
	
	\subsection{Coefficient of the lowest mixed-sign branch cut}
	We now specialize the general result of Sec.~\ref{sec:mix-general-coefficient} to the lowest mixed-sign channel, $00^{+-}$. 
		Let us define 
	\begin{equation}
		\wn_\pm(\qn)=\wn_0^\pm(\qn),
		\qquad
		u_\pm=u_0^\pm,
		\qquad
		D_\pm=D_0^\pm .
	\end{equation}
	At $\kappa=0$, the two ordered assignments $(+,-)$ and $(-,+)$ have the same pair frequency and give equal contributions. It is therefore convenient to define
	\begin{equation}
		\Omega_{00}^{+-}(\mathfrak q)
		=
		\Omega_{00}^{-+}(\mathfrak q)
		\equiv
		\Omega(\mathfrak q)
		=
		\mathfrak w_+(\mathfrak q)
		+
		\mathfrak w_-(\mathfrak q).
	\end{equation}
	Using the mixed-sign thermal weight of Eq.~\eqref{eq:mixed-thermal}, their ordered coefficients satisfy
	\begin{equation}
		T_{00}^{+-}(\mathfrak q)
		=
		T_{00}^{-+}(\mathfrak q)
		\equiv
		T(\mathfrak q)
		=
		\frac12
		\left[
		n_B\!\left(\mathfrak w_+(\mathfrak q)\right)
		-
		n_B\!\left(\mathfrak w_+(\mathfrak q)^*\right)
		\right],
	\end{equation}
	while reflection symmetry \eqref{G_R_bb_symmetry} gives
	\begin{equation}
		D_-(\mathfrak q)
		=
		-\left[D_+(\mathfrak q)\right]^* .
	\end{equation}
	Because multiplication of the two QNM profiles is unchanged when the two internal assignments are exchanged, the corresponding radial vertices are equal:
	\begin{equation}
		v(\mathfrak q)
		\equiv
		v_{00}^{+-}\!\left(\Omega(\mathfrak q),\mathfrak q\right)
		=
		v_{00}^{-+}\!\left(\Omega(\mathfrak q),\mathfrak q\right).
	\end{equation}
	The two ordered mixed-sign contributions therefore add:
	\begin{equation}\label{H_00_mix}
		G_{00}^{\rm mix}
		=
		G_{00}^{+-}
		+
		G_{00}^{-+},
		\qquad
		H_{00}^{\rm mix}(\mathfrak q)
		=
		\frac{
			2T(\mathfrak q)\,v(\mathfrak q)^2
		}{
			D_+(\mathfrak q)D_-(\mathfrak q)
		}.
	\end{equation}
	The remaining question is whether \(T(q)v(q)^2\) is nonzero for the channel of interest. For the lowest mixed-sign channel, we answer this explicitly in Appendix~\ref{app:mixed_numerics} by evaluating the endpoint coefficient and showing that
	$H_* \equiv H_{00}^{\rm mix}(0)\neq0.$
	The resulting contribution is
	\begin{equation}\label{G_00_mix}
	G_{00}^{\rm mix}(\wn)
	=
	\epsilon
	\int_0^\infty
	\frac{\qn^2\,d\qn}{2\pi^2}
	\frac{H_{00}^{\rm mix}(\qn)}
	{\wn-\Omega(\qn)} .
\end{equation}
	
	\section{Numerical construction of the lowest mixed-sign endpoint}
	\label{app:mixed_numerics}
	
	This appendix gives the numerical details underlying the lowest
	mixed-sign endpoint studied in Sec.~\ref{sec:mix-lowest}.  We explain separately the
	determination of the QNM dispersion near $\qn=0$, the radial vertex at the
	endpoint, and the resulting boundary coefficient. 
	
	\subsection{Numerical determination of the endpoint curvature}
	\label{app:mixed_curvature}

	We use the Wronskian \(W_R(\wn,q)\) defined in Eq.~\eqref{eq:retarded-wronskian}. Numerically, \(u_{\rm in}\) is generated from its unit-ingoing Frobenius expansion at the horizon, while \(u_N\) is generated from its unit-normalizable Frobenius expansion at the AdS boundary. The two solutions are evaluated at a common matching point \(z_m\) and inserted into		
	    \begin{equation}	\label{eq:app_WR_numeric}
	    	W_R(\mathfrak w,\mathfrak q)
	    	=
	    	\frac{f(z_m)}{z_m^3}
	    	\left[
	    	u_N(z_m;\mathfrak w,\mathfrak q)
	    	\partial_z u_{\rm in}(z_m;\mathfrak w,\mathfrak q)
	    	-
	    	u_{\rm in}(z_m;\mathfrak w,\mathfrak q)
	    	\partial_z u_N(z_m;\mathfrak w,\mathfrak q)
	    	\right].
	    \end{equation}

For the scalar considered in the main text, the radial equation is $\mathcal{L}_{\wn,\qn}u=0$ with  $\mathcal{L}_{\wn,\qn}$ given by Eq.~\eqref{eq:radial-operator}.
The ingoing solution is constructed near the horizon as
\begin{equation}\label{u_in_exp}
	u_{\rm in}(z;\mathfrak w,\mathfrak q)
	=
	(1-z)^{-i\mathfrak w/4}
	\sum_{n=0}^{N_h}
	a_n^{\rm in}(\mathfrak w,\mathfrak q)(1-z)^n,
	\qquad
	a_0^{\rm in}=1 ,
\end{equation}
while the normalizable solution is constructed near the AdS boundary as
\begin{equation}\label{u_N_exp}
	u_N(z;\wn,\qn)
	=
	z^\Delta
	\sum_{n=0}^{N_N}
	a_n^{N}(\wn,\qn)z^n,
	\qquad
	a_0^{N}=1.
\end{equation}
The coefficients in both series are obtained recursively by substituting
the corresponding Frobenius expansion into the radial equation.	
	
Solving \(W_R(\wn,0)=0\), Eq.~\eqref{eq:retarded-wronskian}, with the numerical solutions \eqref{u_in_exp} and \eqref{u_N_exp}, we find for the fundamental mode with positive real part \footnote{As a check of the numerical implementation, for \(m^2L^2=0\) (\(\Delta=4\)) we reproduce the standard scalar quasinormal frequencies of Ref.~\cite{Nunez:2003eq}. Here, however, we work with $m^2L^2=1$.}
	\begin{equation}
		\wn_+(0)
		=
		3.33918
		-
		2.98394\,i .
		\label{eq:app_omega_plus}
	\end{equation}
	The reflected mode obeys
		$\wn_-(\qn)=-\wn_+(\qn)^* ,$
	and hence
	\begin{equation}
		\wn_*
		=
		\wn_+(0)+\wn_-(0)
		=
		-5.96788\,i .
		\label{eq:app_endpoint}
	\end{equation}
	
To determine the quadratic curvature, i.e., $a_{+-}$ defined in Eq.~\eqref{quadratic_expansion} it,  is useful to introduce
\begin{equation}
	\xi \equiv \qn^2 .
\end{equation}
The QNM equation becomes
\begin{equation}
	W_R\!\left(\mathfrak w_+(\xi),\xi\right)=0 .
\end{equation}
Differentiation at $\xi=0$ gives
\begin{equation}\label{d}
	d\equiv
	\left.\frac{d\mathfrak w_+}{d\xi}\right|_{\xi=0}
	=
	-\left.
	\frac{\partial_\xi W_R}
	{\partial_{\mathfrak w}W_R}
	\right|_{\mathfrak w=\mathfrak w_+(0),\,\xi=0}.
\end{equation}
	We evaluate	the derivatives of $W_R$ with symmetric fourth-order finite
	differences,
\begin{align}\label{par_xi}
	\partial_\xi W_R
	&\simeq
	\frac{
		W_R(\mathfrak w_0,-2h_\xi)
		-8W_R(\mathfrak w_0,-h_\xi)
		+8W_R(\mathfrak w_0,h_\xi)
		-W_R(\mathfrak w_0,2h_\xi)
	}{
		12h_\xi
	},
	\\\label{par_w}
	\partial_{\mathfrak w}W_R
	&\simeq
	\frac{
		W_R(\mathfrak w_0-2h_{\mathfrak w},0)
		-8W_R(\mathfrak w_0-h_{\mathfrak w},0)
		+8W_R(\mathfrak w_0+h_{\mathfrak w},0)
		-W_R(\mathfrak w_0+2h_{\mathfrak w},0)
	}{
		12h_{\mathfrak w}
	},
\end{align}	
The finite-difference formula in Eqs.~\eqref{par_xi} require evaluating \(W_R\) at negative \(\xi\). Since \(\xi=\qn^2\), these points correspond to the analytic continuation \(\qn=i\sqrt{|\xi|}\). Using Eqs.~\eqref{par_xi} and \eqref{par_w} in Eq.~\eqref{d}, we obtain
	\begin{equation}
		d
		=
		0.22384
		+
		0.07321\,i .
		\label{eq:app_d_numeric}
	\end{equation}
	Therefore from Eq.~\eqref{quadratic_expansion}, 
	\begin{equation}
			a_{+-}
			=
			d-d^*
			=
			0.14643\,i .
		\label{eq:app_a_pm}
	\end{equation}
	
	As an independent check, we solve the QNM equation directly at \(\xi=\pm10^{-4}\) and estimate the derivative \(d\mathfrak w_+/d\xi|_{\xi=0}\) using a centered finite difference. This gives
	\begin{equation}
		d_{\rm root}
		=
		0.22384
		+
		0.07321\,i ,
	\end{equation}
	which agrees with Eq.~\eqref{eq:app_d_numeric} to approximately
	$1.5\times10^{-11}$.	
	
	\subsection{Radial vertex at the endpoint}
	\label{app:radial_vertex}

      To determine the endpoint coefficient \(H_*\), we need the radial vertex \(v_*\) defined in Eq.~\eqref{radial_overlap}, evaluated at the lowest mixed-sign endpoint. The QNM pole singularities of the two internal propagators have already been extracted into their residues, so \(v_*\) is the remaining radial overlap and is expected to be finite at the QNM frequencies.
      
      A subtlety arises if the integral in Eq.~\eqref{radial_overlap} is evaluated directly at these frequencies. At the endpoint, both internal QNM wavefunctions and the external retarded bulk-to-boundary propagator are ingoing at the future horizon, \(z\to1\),
      \begin{equation}
      	K_R^\phi(z;\mathfrak w_\star,0)\,
      	u_{\rm in}(z;\mathfrak w_+(0),0)\,
      	u_{\rm in}(z;\mathfrak w_-(0),0)
      	\sim
      	(1-z)^{-\frac{i}{4}
      		[\mathfrak w_*+\mathfrak w_+(0)+\mathfrak w_-(0)]}.
      \end{equation}
       Since
      $ \mathfrak w_*=\mathfrak w_+(0)+\mathfrak w_-(0),$
      this reduces to
      $(1-z)^{-i\mathfrak w_*/2}.$
      For the endpoint considered here,
      $\mathfrak w_*=-5.96788\,i,$
      and hence the integrand behaves as
      $(1-z)^{-2.98394},$
      which makes the radial integral diverge at the horizon.
       We therefore define the radial vertex by first evaluating the integral in a region of complex frequencies where the horizon integral converges, and then analytically continuing the resulting function to the endpoint frequencies.

	The analitically continued radial vertex is then written as
	\begin{equation}
		v_*
		=
		{\rm AC}\int_0^1\frac{dz}{z^5}\,
		K_R^\phi(z;\mathfrak w_*,0)\,
		u_{\rm in}(z;\mathfrak w_+(0),0)\,
		u_{\rm in}(z;\mathfrak w_-(0),0),
	\label{eq:app_vstar}
	\end{equation}
	where $\operatorname{AC}$ denotes the above-mentioned analytic continuation.
	The two internal solutions are normalized to unit ingoing amplitude at
	the horizon.  The external leg is the source-normalized retarded
	bulk-to-boundary propagator \eqref{K_phi_R}.
	
		At the AdS boundary,
\begin{equation}
	K_R^\phi\sim z^{4-\Delta},
	\qquad
	u_{\rm in}(z;\mathfrak w_\pm,0)
	\sim B_\pm z^\Delta ,
\end{equation}
	so that
\begin{equation}
	z^{-5}
	K_R^\phi
	u_{\rm in}(z;\mathfrak w_+,0)
	u_{\rm in}(z;\mathfrak w_-,0)
	\sim z^{\Delta-1}.
\end{equation}
	and the radial integral is convergent at $z=0$.
	
	As mentioned above, the horizon requires more care.  Let us set
		$x=1-z .$
	The three radial functions have Frobenius expansions about $x=0$
\begin{align}
	u_{\rm in}(z;\mathfrak w_+,0)
	&=
	x^{-i\mathfrak w_+/4}
	\sum_{n=0}^{\infty}a_n^{(+)}x^n,
	\\
	u_{\rm in}(z;\mathfrak w_-,0)
	&=
	x^{-i\mathfrak w_-/4}
	\sum_{n=0}^{\infty}a_n^{(-)}x^n,
	\\
	K_R^\phi(z;\mathfrak w_*,0)
	&=
	\frac{x^{-i\mathfrak w_*/4}}
	{A(\mathfrak w_*,0)}
	\sum_{n=0}^{\infty}a_n^{(*)}x^n.
\end{align}	In addition,
	\begin{equation}
		z^{-5}
		=
		(1-x)^{-5}
		=
		\sum_{r=0}^{\infty}
		\binom{r+4}{4}x^r .
	\end{equation}
		Multiplying these series gives
\begin{equation}
	\frac{1}{z^5}
	K_R^\phi
	u_{\rm in}(z;\mathfrak w_+,0)
	u_{\rm in}(z;\mathfrak w_-,0)
	=
	x^p\sum_{j=0}^{\infty}\mathfrak c_jx^j,
	\label{eq:app_integrand_series}
	\end{equation}
	where
\begin{equation}
	p
	=
	-\frac{i}{4}
	\left(
	\mathfrak w_*+\mathfrak w_+ +\mathfrak w_-
	\right)
	=
	-\frac{i\mathfrak w_*}{2},
	\label{eq:app_p}
	\end{equation}
	and
\begin{equation}	\label{eq:app_cj}
		\boxed{
\mathfrak 	c_j
	=
	\frac{1}{A(\mathfrak w_*,0)}
	\sum_{r+s+t+u=j}
	\binom{r+4}{4}
	a_s^{(*)}
	a_t^{(+)}
	a_u^{(-)} .
	}
	\end{equation}
	This shows how the coefficients entering the horizon continuation, Eq.~\eqref{eq:app_integrand_series}, are obtained directly from the Frobenius coefficients of
	the three radial solutions.
	
	For $\text{Re}\, p>-1$ the near-horizon integral is convergent and gives
	\begin{equation}
		\int_0^{x_h} dx\,
		x^p
		\sum_{j=0}^{\infty}\mathfrak c_jx^j
		=
		\sum_{j=0}^{\infty}
		\frac{
			\mathfrak c_j x_h^{p+j+1}
		}{
			p+j+1
		}.
		\label{eq:app_AC}
	\end{equation}
For the QNM frequencies relevant here, Eq.~\eqref{eq:app_p} gives \(p=-2.983940057\ldots\), so the integral on the left-hand side of Eq.~\eqref{eq:app_AC} is divergent as an ordinary real-\(x\) integral. The expression on the right-hand side, however, agrees with that integral in the convergence domain \(\mathrm{Re}\,p>-1\) and provides its meromorphic continuation in \(p\). Since \(p\) is not a negative integer at the QNM point, none of the denominators \(p+j+1\) vanishes. The continued radial integral is therefore finite there.

	Numerically, we split the radial interval into a directly integrated
	part and a near-horizon part,
	\begin{equation}
		v_*
		=
		v_{\rm dir}(x_h)
		+
		v_{\rm hor}^{\rm AC}(x_h),
		\label{eq:app_vertex_split}
	\end{equation}
	where the second term is evaluated using Eq.~\eqref{eq:app_AC}. The two terms separately depend strongly on \(x_h\), while their sum does not. As a numerical stability check, we vary both the split point \(x_h\) and the truncation order of the near-horizon Frobenius series. This gives
	\begin{equation}
	\boxed{v_\star=1.173928.}
	\end{equation}
	Over the tested range of \(x_h\) and Frobenius truncation orders, the result varies by less than approximately \(6\times10^{-14}\).

	\subsection{QNM residue, thermal weight, and endpoint coefficient}
	\label{app_QNM_residue}
	
	The radial solutions constructed in Sec~\ref{app:mixed_curvature} also determine the QNM residue with respect to
	the dimensionless frequency $\mathfrak w$.  From Eq.~\eqref{eq:qnm-profile}, at the fundamental QNM, we extract the proportionality factor at the matching point,
	\begin{equation}
		B_+
		=
		\frac{
			u_{\rm in}\!\left(z_m;\mathfrak w_+(0),0\right)
		}{
			u_N\!\left(z_m;\mathfrak w_+(0),0\right)
		}
		=
		3.202891
		+
		0.290271\,i .
		\label{eq:app_Bplus}
	\end{equation}
	The result is unchanged to the displayed precision when the matching point $z_m$ is varied.

Using the same numerical construction as in  Sec.~\ref{app:mixed_curvature},  We find
\begin{equation}
	\left.
	\partial_{\mathfrak w}A(\mathfrak w,0)
	\right|_{\mathfrak w=\mathfrak w_+(0)}
	=
	0.047995
	+
	0.070824\,i .
\end{equation}
Equation~\eqref{eq:qnm-residue} then gives
\begin{equation}
D_+
=
-2\nu B_+
\left.
\partial_{\mathfrak w}A(\mathfrak w,0)
\right|_{\mathfrak w=\mathfrak w_+(0)}
=
-0.595535-1.076777\,i .
\end{equation}
Reflection symmetry implies
$D_-=-D_+^*
=
0.595535-1.076777\,i .$

The thermal factor requires no further radial calculation. For the lowest mixed-sign channel,
\begin{equation}
	T_*=\frac12
	\left[
	n_B(\mathfrak w_+)-n_B(\mathfrak w_+^*)
	\right]	=
	1.402096\times10^{-6}\,i .
	\label{eq:app_Tstar}
\end{equation}
The remaining ingredient is the analytically continued radial vertex $v_*$
computed in Sec.~\ref{app:radial_vertex}, given by Eq.~\eqref{eq:app_vstar}. Combining these ingredients gives
\begin{equation}
	H_*
	=
	\frac{2T_*v_*^2}{D_+D_-}
	=
	-5.104611\times10^{-6}\,i .
	\label{eq:app_Hstar}
\end{equation}

    \section{Complete Schwinger--Keldysh loop and numerical continuation}
    \label{app:fullSK}
      In this appendix we describe the complete Schwinger--Keldysh loop used in
   the numerical test of Sec.~\ref{sec:mix-lowest}.
   The one-loop retarded self-energy is given in Eq.~\eqref{eq:self-energy-fourier}.  Using the KMS
   relation Eq.~\eqref{eq:symmetric-propagator}, and suppressing radial arguments, we write at fixed
   $\mathfrak q$
  \[
  R(\nu,\mathfrak q)\equiv G_R^{bb}(\nu,\mathfrak q),
  \qquad
  A(\nu,\mathfrak q)\equiv G_A^{bb}(\nu,\mathfrak q).
  \]


Using the KMS relation to rewrite the symmetric propagators in Eq.~\eqref{eq:self-energy-fourier}, we collect the complete integrand of the internal-frequency convolution at fixed \(\mathfrak q,z,z'\) into the kernel
\begin{equation}	\label{eq:app_complete_SK_kernel_expanded}
		\mathcal K
		=
		\frac{i}{2}\,
		c(\mathfrak w-\mathfrak w')
		R(\mathfrak w',\mathfrak q)
		\left[
		R(\mathfrak w-\mathfrak w',\mathfrak q)
		-
		A(\mathfrak w-\mathfrak w',\mathfrak q)
		\right],
\end{equation}
where we have suppressed the radial arguments $z,z'$. 
The retarded bubble in the upper half-plane is then
\begin{equation}
	\frac{\Sigma_R(\wn;z,z')}{\epsilon}
	=
	\int_0^\infty\frac{\qn^2d\qn}{2\pi^2}
	\int_{C_0}\frac{d\wn'}{2\pi}\,
	{\cal K}(\wn';\wn,\qn; z, z'),
	\qquad
	C_0=\mathbb R,
	\qquad
	\operatorname{Im}\wn>0 .
	\label{eq:app_SK_contour_start}
\end{equation}

We now analytically continue the function defined by
Eq.~\eqref{eq:app_SK_contour_start}.  Let $\wn$ move continuously
along a path $\gamma$ beginning in the upper half-plane.  The
singularities of ${\cal K}$ then move in the complex $\wn'$ plane.
For example, the two retarded factors in the first term of
Eq.~\eqref{eq:app_complete_SK_kernel_expanded} have poles at
\begin{equation}
	\wn'=\wn_m^\sigma(\qn),
	\qquad
	\wn'=\wn-\wn_n^\tau(\qn),
	\label{eq:app_moving_RR_poles}
\end{equation}
while the advanced factors give the corresponding reflected pole
families.  The thermal functions $c(\wn')$ and
$c(\wn-\wn')$ supply their own Matsubara poles.

The analytic continuation of the retarded loop is obtained by
continuously transporting the initial contour $C_0$ so that no
singularity crosses it.  Thus a pole that lies above $C_0$ at the
starting point remains above the transported contour, and similarly for
a pole that starts below it (see Fig.~\ref{fig:mix-frequency-pinch}).  We denote the resulting contour by
$C_\gamma$.

In the numerical calculation we do not explicitly parametrize
$C_\gamma$ at every point of the continuation.  Instead, the complete
kernel~\eqref{eq:app_complete_SK_kernel_expanded} is integrated numerically on a
fixed reference contour, which is taken to be the real $\wn'$ axis here, while all singularities are tracked as
$\wn$ is varied.  Whenever the transported contour and the reference
contour pass on different sides of a pole, their difference is included
through the residue of the \emph{complete} kernel.  Schematically,
\begin{equation}
	\int_{C_\gamma}\frac{d\wn'}{2\pi}\,
	{\cal K}(\wn')
	=
	\int_{C_{\rm ref}}\frac{d\wn'}{2\pi}\,
	{\cal K}(\wn')
	+
	i\sum_{p\in{\cal P}_\gamma}
	s_p\,
	\underset{\wn'=p}{\operatorname{Res}}
	{\cal K}(\wn'),
	\label{eq:app_contour_transport}
\end{equation}
where ${\cal P}_\gamma$ is the set of poles lying between the two
contours and $s_p=\pm1$ records the orientation.  The residue in
Eq.~\eqref{eq:app_contour_transport} is therefore not merely a
single-propagator residue: all other propagators, thermal factors and
radial dependence in ${\cal K}$ are retained and evaluated at the pole.

\begin{figure}[t]
	\centering
	\includegraphics[
	width=1.\textwidth,
	trim=0 2 0 0,
	clip
	]{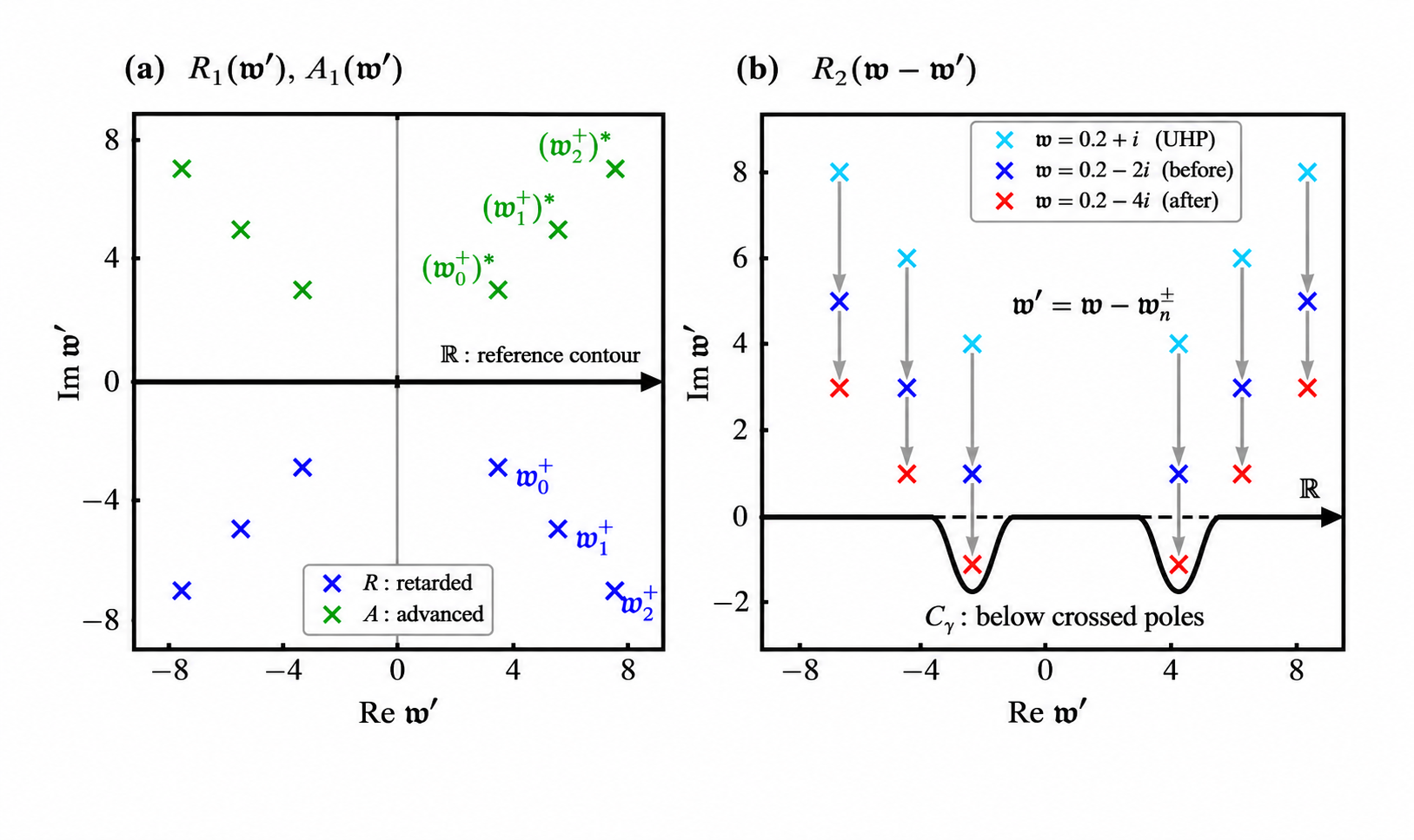}
\caption{
	Pole structure used in the numerical analytic continuation. \textbf{Left:} retarded and advanced bulk-propagator poles in the complex \(\mathfrak w'\) plane at fixed loop momentum. The real axis is the reference contour \(\mathbb R\). \textbf{Right:} poles of the shifted retarded propagator \(R_2(\mathfrak w-\mathfrak w')\) as the external frequency \(\mathfrak w\) is continued downward. The three displayed values of \(\mathfrak w\) lie on the vertical part of the path \(\gamma_+\) shown in Fig.~\ref{fig:gamma_paths}. When a shifted pole crosses the real reference contour, the analytically transported contour must pass below it in order to preserve its relative position with respect to the pole. In the numerical implementation, we keep the reference contour fixed and account for the corresponding difference by adding the residue of the complete Schwinger–Keldysh kernel, Eq.~\eqref{eq:app_contour_transport}. We omit the thermal Matsubara poles for clarity. We also suppress the radial arguments, since they affect the pole residues but not the QNM pole locations.
}
\label{fig:R_1_A_1}
\end{figure}

Figure~\ref{fig:R_1_A_1} illustrates this construction.
At the starting point in the retarded domain,
$\operatorname{Im}\mathfrak w>0$, the poles of
$R_1(\mathfrak w')$ lie below the real $\mathfrak w'$ axis, whereas
the poles of the shifted factor
$R_2(\mathfrak w-\mathfrak w')$ lie above it.  As the external
frequency $\mathfrak w$ is continued downward, the latter poles move
downward and can cross the real reference contour.  The analytically
transported contour must then pass below the crossed poles so that each
pole remains on the same side of the contour as at the starting point.
Numerically, we keep the reference contour fixed and account for this
difference through the complete kernel residues in Eq.~\eqref{eq:app_contour_transport}.

We evaluate the regular part of the frequency integral numerically. When a pole approaches the integration contour, direct numerical integration becomes inaccurate because the integrand varies rapidly near the pole. We therefore subtract the local pole term \(r_p/(\wn'-p)\), integrate the remaining smooth function numerically, and add back the pole term using its known analytic integral.  Higher-order thermal poles are handled in the same way using
their Laurent expansion.  At finite frequency cutoff, we also include the short endpoint segments required to complete the contour deformation.

The precise geometric shape of the external-frequency path $\gamma$ is
not part of the continuation prescription.  Any continuous deformation of
$\gamma$ that keeps its endpoints fixed and does not cross a singularity
defines the same analytic continuation.  In the numerical calculation we
choose a simple representative: the two paths start at
$\mathfrak w=\pm0.2+i$, descend at fixed real part, and then approach the
common point $\mathfrak w=-is$ laterally from the right and left,
respectively.  These two choices define the two sides of the cut. See Fig.~\ref{fig:gamma_paths}.

For the corresponding loop-momentum continuation we use
\begin{equation}
	\mathfrak q_\pm(u)
	=
	u\pm iu_0\sin\!\left(\frac{\pi u}{Q}\right).
	\qquad
	0\le u\le Q ,
\end{equation}
Here \(Q\) is the finite momentum cutoff used in the numerical integral, while \(u_0\) controls the height of the deformation away from the real \(\mathfrak q\) axis. Neither parameter specifies the analytic continuation itself: \(Q\) is taken sufficiently large for cutoff stability, and \(u_0\) may be varied continuously provided that no singularity crosses the contour. The two paths have the same endpoints, \(\qn_\pm(0)=0\) and \(\qn_\pm(Q)=Q\), with \(\qn_+\) deforming into the upper-half \(\qn\) plane and \(\qn_-\) into the lower-half \(\qn\) plane.

	\section{Large-momentum QNM scaling, residues, and radial vertices}
	\label{app:largeq}
This appendix derives the large-\(\mathfrak q\) behavior needed for the accumulation and late-time analysis of Secs.~\ref{sec:origin-accumulation} and \ref{sec:real-time}. We begin with the large-\(\mathfrak q\) QNM spectrum of Ref.~\cite{Festuccia:2008zx}, specialized to the bulk scalar considered here. We then determine the associated asymptotics of the QNM residues and radial vertices that enter the mixed-sign branch-cut weight. Throughout, the overtone indices are kept fixed while $\mathfrak q\to\infty$.
	
	\subsection{Large-\(\mathfrak q\) QNM asymptotics}
	\label{app:largeq-spectrum}
	
	For the scalar considered in the main text,
	\begin{equation}
		m^2=1,
		\qquad
		\nu=\sqrt{4+m^2}=\sqrt5 .
	\end{equation}
	For fixed overtone index $n$ and $\mathfrak q\to\infty$, the relevant
	radial region is the boundary layer
	$z\sim\mathfrak q^{-1/3}$~\cite{Festuccia:2008zx}.  We therefore introduce
	\begin{equation}
		y=\mathfrak q^{1/3}z,
		\qquad
		\mathfrak w_n^+(\mathfrak q)
		=
		\mathfrak q+c_n\mathfrak q^{-1/3}
		+O(\mathfrak q^{-5/3}) .
	\end{equation}
	Expanding the radial Klein--Gordon equation at fixed $y$ gives, to leading
	order,
	\begin{equation}
		U_n''-\frac{3}{y}U_n'
		+
		\left(
		y^4+2c_n-\frac{m^2}{y^2}
		\right)U_n=0 .
	\end{equation}
	With
	\begin{equation}\label{U_n}
		U_n(y)=y^{3/2}\psi_n(y),
	\end{equation}
	this becomes
	\begin{equation}\label{app:eq:quartic}
		\psi_n''
		+
		\left[
		y^4+2c_n-\frac{\nu^2-\frac14}{y^2}
		\right]\psi_n=0,
		\qquad
		\psi_n(y)\sim y^{\nu+1/2}
		\quad (y\to0).
	\end{equation}
	Although $0\le z\le1$, the scaling coordinate $y=\qn^{1/3}z$ has upper endpoint $y_{\max}=\qn^{1/3}$, which tends to infinity as $\qn\to\infty$. To match the boundary-layer solution to the solution outside the layer, we require $z\gg \qn^{-1/3}$, or \(y\gg1\), while remaining in the near-boundary region $z\ll1$, or $y\ll \qn^{1/3}$. The overlap region is therefore
	\begin{equation}
	1\ll y\ll \qn^{1/3}
	\qquad
	(\qn^{-1/3}\ll z\ll1).
	\end{equation}
	In the large-$q$ scaling problem this overlap is represented by the large-\(y\) asymptotic limit.
		
	 In the matching region,
	the scaling solution that matches onto the ingoing QNM has the asymptotic behavior
	\begin{equation}\label{large_y}
		\psi_n(y)\sim e^{iy^3/3},
		\qquad
		y\to\infty .
	\end{equation}
	Rotating to the decaying ray,
	\begin{equation}
		y=e^{i\pi/6}x,
	\end{equation}
	maps Eq.~\eqref{app:eq:quartic} to the real half-line eigenvalue problem
	\begin{equation}\label{phi_lambda}
		\left[
		-\frac{d^2}{dx^2}
		+x^4
		+\frac{\nu^2-\frac14}{x^2}
		\right]\varphi_n(x)
		=
		\lambda_n\varphi_n(x),\,\,\,\,\,\,\,\,\,\,\,
		0\le x<\infty,
	\end{equation}
	with
	\begin{equation}\label{bdy}
		\varphi_n(x)\sim x^{\nu+1/2}
		\quad (x\to0),
		\qquad
		\varphi_n(x)\to0
		\quad (x\to\infty).
	\end{equation}
	With the boundary conditions in Eq.~\eqref{bdy}, $\mathcal H_\nu
	\equiv
	-\frac{d^2}{dx^2}
	+x^4+\frac{\nu^2-\tfrac14}{x^2}$ is positive and confining, so its spectrum is discrete and ordered as
	\begin{equation}
	0<\lambda_0<\lambda_1<\lambda_2<\cdots .
	\end{equation}
	Comparison with Eq.~\eqref{app:eq:quartic} gives
	\begin{equation}
		2c_n=e^{-i\pi/3}\lambda_n .
	\end{equation}
	Thus, for every fixed overtone index, we obtain Eq.~\eqref{eq:largeq-qnm}.
		For the lowest mode, one finds
	\begin{equation}
		\lambda_0=9.820672149995,
		\qquad
		a_0
		\equiv
		\frac{\sqrt3}{2}\lambda_0
		=
		8.504951564134.
	\end{equation}

	\subsection{QNM normalization and pole residues}
	\label{App_QNM_ren}
	
	The QNM residue is given in \eqref{eq:qnm-residue}.
	To determine its large-$\mathfrak q$ behavior, we first rewrite $D_n$
	as a radial integral. For this purpose let us define
	\[
	u_n\equiv u_{\text{in}}(z;\mathfrak w_n^+,\mathfrak q),\,\,\,\,\,\,\,\,\,v_n(z)\equiv
	\left.
	\partial_{\mathfrak w}
	u_{\rm in}(z;\mathfrak w,\mathfrak q)
	\right|_{\mathfrak w=\mathfrak w_n^+}.
	\]
	Differentiating the radial equation
	$\mathcal L_{\mathfrak w,\mathfrak q}u_{\rm in}=0$, with $\mathcal L_{\mathfrak w,\mathfrak q}$ given by Eq.~\eqref{eq:radial-operator}, with respect to \(\mathfrak w\), and then evaluating at
	\(\mathfrak w=\mathfrak w_n^+(\mathfrak q)\), gives
	\begin{equation}
		\mathcal L_n v_n
		=
		-\frac{2\mathfrak w_n^+}{z^3f(z)}\,u_n,
		\qquad
		\mathcal L_n u_n=0 .
	\end{equation}
	Multiplying the first equation by $u_n$, subtracting
	$v_n\mathcal L_nu_n$, and collecting the derivative terms yields
	\begin{equation}\label{eigen_value_EQ}
		\partial_z
		\left[
		\frac{f}{z^3}
		\left(
		u_n\partial_zv_n-v_n\partial_zu_n
		\right)
		\right]
		=
		-\frac{2\mathfrak w_n^+}{z^3f}\,u_n^2 .
	\end{equation}
	On th other hand, 
	near the AdS boundary,
	\[
	u_n(z)=B_n z^\Delta+\cdots,
	\qquad
	v_n(z)=
	\partial_{\mathfrak w}A(\mathfrak w_n^+,\mathfrak q)
	\,z^{4-\Delta}+\cdots ,
	\]
	so that
	\begin{equation}
		\left.
		\frac{f}{z^3}
		\left(
		u_n\partial_zv_n-v_n\partial_zu_n
		\right)
		\right|_{z\to0}
		=
		-2\nu B_n\partial_{\mathfrak w}A
		=
		D_n .
	\end{equation}
	Evaluating Eq.~\eqref{eigen_value_EQ} first in the convergence domain, where the
	continued QNM solution gives no horizon endpoint contribution, and then
	analytically continuing to the QNM frequency, as in Sec.~\ref{app:radial_vertex},  gives
	\begin{equation}
		\boxed{
			D_n(\mathfrak q)
			=
			\operatorname{AC}\int_0^1 dz\,
			\frac{2\mathfrak w_n^+(\mathfrak q)}
			{z^3f(z)}
			\,u_n(z)^2 .
		}
	\end{equation}
	We now evaluate this expression in the large-$\mathfrak q$ boundary layer.
	
	In the boundary layer the full QNM solution and the scaling solution of Sec.~\ref{app:largeq-spectrum} obey the same leading equation but have different normalizations. We therefore write
	\begin{equation}\label{u_n_U_n}
	u_n(z)=\mathcal C_n(\mathfrak q)U_n(y),
	\qquad y=\mathfrak q^{1/3}z.
	\end{equation}
	Since \(u_n\sim B_nz^\Delta\) while  \(U_n(y)\sim y^\Delta\)\footnote{This directly follows from Eqs.~\eqref{U_n} and \eqref{app:eq:quartic}.}, one has
	$\mathcal C_n=B_n\mathfrak q^{-\Delta/3}.$
	Using $\mathfrak w_n^+\sim\mathfrak q$ and $f\to1$, we obtain
	\[
	\frac{2\mathfrak w_n^+\,dz}{z^3f}
	\sim
	2\mathfrak q^{5/3}\frac{dy}{y^3}.
	\]
	Hence
	\begin{equation}\label{D_n_N_n}
		D_n(\mathfrak q)
		\sim
		2\mathfrak q^{5/3}C_n^2 N_n,
		\qquad
		N_n\equiv
		\int_{e^{i\pi/6}\mathbb R_+}dy\,\psi_n(y)^2,
	\end{equation}
	where we have used Eq.~\eqref{U_n}. Now by using the rotation $y=e^{i\pi/6}x$ of Sec.~\ref{app:largeq-spectrum}, we find
	\begin{equation}\label{N_n}
		N_n
		=
		e^{i\pi(\nu+1)/3}
		\int_0^\infty dx\,\varphi_n(x)^2 .
	\end{equation}
	
	Substituting $u_n$, from Eq.~\eqref{u_n_U_n}, and $D_n(\qn)$, from Eq.~\eqref{D_n_N_n}, into Eq.~\eqref{eq:qnm-residue}, the unknown matching normalization $\mathcal C_n$ cancels between the numerator
	and denominator of the residue, leaving
	\begin{equation}
		\boxed{
			r_{n,+}(\mathfrak q;z,z')
			\sim
			\frac{U_n(y)U_n(y')}
			{2\mathfrak q^{5/3}N_n},
			\qquad
			y=\mathfrak q^{1/3}z,
			\quad
			y'=\mathfrak q^{1/3}z' .
		}
	\end{equation}
	Note that for real $\mathfrak q,z,z'$, the reflected pole obeys
		$r_{n,-}(\mathfrak q;z,z')
		=
		-r_{n,+}(\mathfrak q;z,z')^* ,$
		as found in Eq.~\eqref{eq:residue-reality}.
	
\subsection{Large-$\mathfrak q$ radial vertex}
	\label{app:radial-overlap}
	
For the mixed-sign channels considered here, the external frequency
approaches the origin as $\mathfrak q\to\infty$.  In the boundary layer
$z\sim\mathfrak q^{-1/3}$, the source-normalized external leg therefore
has its universal near-boundary form (see 	Eqs.~\eqref{K_phi_R}, 	\eqref{eq:source-response}, and	\eqref{eq:boundary-solutions})
\begin{equation}
	K_R(\mathfrak w,0;z)\sim z^{2-\nu}.
\end{equation}

For a diagonal mixed-sign pair, the two internal scaling profiles are
reflected partners, so that on real $y$
\begin{equation}
U_{n,+}(y)U_{n,-}(y)=|U_n(y)|^2
=y^3|\psi_n(y)|^2 .
\end{equation}
Using $z=\mathfrak q^{-1/3}y$, the radial vertex therefore contains
\begin{align}
	\int\frac{dz}{z^5}\,
	K_R\,|U_n|^2
	&\sim
	\mathfrak q^{(\nu+2)/3}
	\int_0^\infty dy\,y^{-\nu}|\psi_n(y)|^2
	\nonumber\\
	&\equiv
	\mathfrak q^{(\nu+2)/3}J_n .
\end{align}
The integral converges at both endpoints.

For the lowest mode we define
\begin{equation}
	\mathcal R\equiv\frac{J_0}{|N_0|},
\end{equation}
where $N_0$ is the $n=0$ case of the general residue normalization given by Eq.~\eqref{D_n_N_n}.
Numerically we find
\begin{equation}
	\boxed{\mathcal R\simeq18.0823752.}
\end{equation}

	\subsection{Lowest mixed-sign radial coefficient}
		\label{app:H00}
	We now combine the residue scaling of Sec.~\ref{App_QNM_ren} with the radial vertex of
	Sec.~\ref{app:radial-overlap}.  For the lowest reflected pair, the product of the two QNM
	residues contributes
	\begin{equation}
		-\frac{1}
		{4\mathfrak q^{10/3}|N_0|^2},
	\end{equation}
	while the two radial vertices contribute
	\begin{equation}
		\mathfrak q^{2(\nu+2)/3}J_0^2 .
	\end{equation}
	Therefore
	\begin{equation}
		\boxed{
			\frac{v(\mathfrak q)^2}
			{D_+(\mathfrak q)D_-(\mathfrak q)}
			\sim
			-\frac{\mathcal R^2}{4}\,
			\mathfrak q^{(2\nu-6)/3},
			\qquad
			\mathcal R=\frac{J_0}{|N_0|}.
		}
	\end{equation}
	This is the radial coefficient used in Eq.~\eqref{H_00_mix_box}.	
	
\subsection{General fixed-index mixed-sign channel}

The preceding counting extends directly to any fixed pair $(m,n)$.
Using the normalization $N_n$ defined in Sec.~\ref{App_QNM_ren}, we define the
radial-vertex coefficient
\begin{equation}
	J_{mn}
	\equiv
	\int_0^\infty dy\,
	y^{-\nu}\psi_m(y)\psi_n(y)^* .
\end{equation}
For the present value of $\nu$ this integral converges for every fixed
pair.  For $m=n$, $J_{nn}$ is real and positive, whereas for
$m\neq n$ it is generally complex and may in principle vanish.

For one ordered $(+,-)$ assignment, the product of the two QNM residues
gives
\begin{equation}
	-\frac{1}
	{4\mathfrak q^{10/3}N_mN_n^*},
\end{equation}
while the two radial vertices give
\begin{equation}
	\mathfrak q^{2(\nu+2)/3}J_{mn}^2 .
\end{equation}
Using  $T_{mn}^{+-}
=\frac{1}{2}\big[n_B(\mathfrak w_m^+)-n_B[(\mathfrak w_n^+)^*\big]
\sim
-\frac{\pi d_{mn}}{2}\,\mathfrak q^{-1/3}e^{-\pi\mathfrak q}$,
Eq.~\eqref{eq:H-general} gives
\begin{equation}
	\boxed{
		H_{mn}^{+-}(\mathfrak q)
		\sim
		B_{mn}^{+-}\,
		\mathfrak q^{(2\nu-7)/3}
		e^{-\pi\mathfrak q},
		\qquad
		B_{mn}^{+-}
		=
		\frac{\pi d_{mn}}{8}
		\frac{J_{mn}^2}{N_mN_n^*}.
	}
\end{equation}
Including the two ordered contributions gives
\begin{equation}
	H_{mn}^{\rm mix}=2H_{mn}^{+-},
	\qquad
	B_{mn}=2B_{mn}^{+-}.
\end{equation}
Thus every mixed-sign channel with $J_{mn}\neq0$ has the common
large-$\mathfrak q$ scaling
\begin{equation}
	H_{mn}^{\rm mix}(\mathfrak q)
	\propto
	\mathfrak q^{(2\nu-7)/3}e^{-\pi\mathfrak q}.
\end{equation}
If $J_{mn}=0$ for a particular off-diagonal pair, the leading term
vanishes and the subleading large-$\mathfrak q$ behavior must be used.

\section{Ordering of the fixed-channel suppression coefficients}
	\label{app:fixed-channel-ordering}
	
	For a mixed-sign pair, we write
	\begin{equation}
	\kappa_{mn}=|\kappa_{mn}|e^{i\theta_{mn}},
	\qquad
	|\theta_{mn}|<\frac{\pi}{6},
	\end{equation}
	with
	$\kappa_{mn}
	=
	\frac14\left[
	\sqrt3(\lambda_m+\lambda_n)
	+i(\lambda_m-\lambda_n)
	\right].$
	For $m=n$, $\kappa_{nn}>0$, and we take the ordinary positive value of $\kappa_{nn}^{3/4}$. For $m\neq n$, $\kappa_{mn}$ moves into the complex plane but remains within $|\theta_{mn}|<\pi/6$. We define $\kappa_{mn}^{3/4}$ by continuously following this positive-real value,
	\begin{equation}
	\kappa_{mn}^{3/4}
	=
	|\kappa_{mn}|^{3/4}e^{3i\theta_{mn}/4}.
	\end{equation}
     Let us reacll that
	\begin{equation}
	\operatorname{Re}\kappa_{mn}
	=
	\frac{\sqrt3}{4}(\lambda_m+\lambda_n)
	\ge
	\frac{\sqrt3}{2}\lambda_0
	=
	\kappa_{00}.
	\end{equation}
		Using
	$C_{mn}
	=
	\frac{4\beta^{1/4}}{3^{3/4}}
	\kappa_{mn}^{3/4},$
	we obtain
	\begin{equation}
	\operatorname{Re}\kappa_{mn}^{3/4}
	=
	|\kappa_{mn}|^{3/4}
	\cos\frac{3\theta_{mn}}4.
	\end{equation}
	For $|\theta_{mn}|<\pi/6$,
	\[
	\cos\frac{3\theta_{mn}}4
	\ge
	(\cos\theta_{mn})^{3/4},
	\]
	and hence
	$\operatorname{Re}\kappa_{mn}^{3/4}
	\ge
	(\operatorname{Re}\kappa_{mn})^{3/4}
	\ge
	\kappa_{00}^{3/4}.$
	Therefore
	\begin{equation}
	\boxed{
		\operatorname{Re}C_{mn}\ge C_{00}.
	}
	\end{equation}
	Equality requires $\theta_{mn}=0$ and
	$\lambda_m+\lambda_n=2\lambda_0$, which together imply
	$m=n=0$.

\end{document}